\documentclass[a4paper,fleqn,usenatbib]{mnras}

\usepackage{newtxtext,newtxmath}

\usepackage[T1]{fontenc}
\usepackage{ae,aecompl}

\usepackage{graphicx}	
\usepackage{amsmath}	
\usepackage{amssymb}	
\usepackage{subfigure}
\usepackage{multirow}
\usepackage{ulem}
\usepackage{url}
\usepackage{lipsum}

\newcommand\blfootnote[1]{%
	\begingroup
	\renewcommand\thefootnote{}\footnote{#1}%
	\addtocounter{footnote}{-1}%
	\endgroup
}

\title[RFI-Net for RFI detection]{Deep residual detection of Radio Frequency Interference for FAST}

\author[Zhicheng Yang et al.]{
	Zhicheng Yang,$^{1}$
	Ce Yu,$^{1}$\thanks{Corresponding Author: Jian Xiao (xiaojian@tju.edu.cn) \newline
		$\textcolor{white}{\star}$\ \ Corresponding Author: Ce Yu (yuce@tju.edu.cn)}
	Jian Xiao,$^{1}\textcolor{blue}{^{\star}}$
	and Bo Zhang$^{2}$$^{3}$
	\\
	$^{1}$College of Intelligence and Computing, Tianjin University, No.135 Yaguan Road, Haihe Education Park, Tianjin, 300350, China\\
	$^{2}$National Astronomical Observatories, Chinese Academy of Sciences, No.20 Datun Road, Chaoyang District, Beijing, 100012, China\\
	$^{3}$CAS Key Laboratory of FAST, National Astronomical Observatories, Chinese Academy of Sciences
}

\date{Accepted 2019 December 13. Received 2019 November 18; in original form 2019 June 18}

\pubyear{2020}

\begin{document}
	\label{firstpage}
	\pagerange{\pageref{firstpage}--\pageref{lastpage}}
	\maketitle
	
	\begin{abstract}
		Radio frequency interference (RFI) detection and excision is one of the key steps in the data processing pipeline of the Five-hundred-meter Aperture Spherical radio Telescope (FAST). The FAST telescope, due to its high sensitivity and large data rate, requires more accurate and efficient RFI flagging methods than its counterparts. In the last decades, approaches based upon artificial intelligence (AI), such as codes using Convolutional Neural Network (CNN), have been proposed to identify RFI more reliably and efficiently. However, RFI flagging of FAST data with such methods has often proved to be erroneous, with further manual inspections required. In addition, network construction as well as training dataset preparation for effective RFI flagging has imposed significant additional workloads. Therefore, rapid deployment and adjustment of AI approaches for different observations is impractical to implement with existing algorithms. To overcome such problems, we propose a model named RFI-Net. With the input of raw data without any processing, RFI-Net can detect RFI automatically, producing corresponding masks without any alteration of the original data. Experiments with RFI-Net using simulated astronomical data show that our model has outperformed existing methods in terms of both precision and recall. Besides, compared with other models, our method can obtain the same relative accuracy with less training data, thus saving effort and time required to prepare the training set. Further, the training process of RFI-Net can be accelerated, with overfittings being minimised, compared with other CNN codes. The performance of RFI-Net has also been evaluated with observing data obtained by FAST and Bleien Observatory. Our results demonstrate the ability of RFI-Net to accurately identify RFI with fine-grained, high-precision masks that required no further modification.
	\end{abstract}
	
	
	
	
	\section{Introduction}
	\label{sec_1_introduction}
	
	Any undesired signal received by radio telescopes can be referred to as radio frequency interference (RFI, see \citet{2017MS&E..198a2012M}). \blfootnote{\textit{This is a pre-copyedited, author-produced PDF of an article accepted for publication in} Monthly Notices of the Royal Astronomical Society \textit{following peer review. The version of record} Zhicheng Yang, Ce Yu, Jian Xiao, Bo Zhang, Deep residual detection of radio frequency interference for FAST, Monthly Notices of the Royal Astronomical Society, Volume 492, Issue 1, February 2020, Pages 1421–1431, https://doi.org/10.1093/mnras/stz3521 \textit{is available online at:} \url{https://doi.org/10.1093/mnras/stz3521}.}The spectral coverage of radio astronomical observations often overlaps with radio emissions originating from modern civilisation. These sources of RFI negatively impact on the data analysis. They can be roughly classified into sporadic emissions, which may erupt occasionally with pulse-like structures, as well as persisting ones (e.g., see \citet{Offringa2010a} and \citet{AnTao2017}). Construction equipment with electric motors, digital cameras, and other similar electronic devices all generate RFI of the former type, usually contaminating multiple channels in the frequency domain (wide band). TV, mobile phone infrastructure as well as radio stations on the other hand usually operate in designated bands. Their emissions, along with harmonics of their local oscillator frequencies, can give rise to the latter persisting narrow-band RFI. All RFI can either originate from devices directly related to radio telescopes themselves (including on-site electronics, network systems, data processing computers, etc.), or come from mobile or fixed sources outside the observatory. Besides, natural radio emitters like the ground, lightning, and the sun can also give off extra RFI \citep{7833529}.
	
	\begin{figure}
		\centering
		\includegraphics[width=\linewidth]{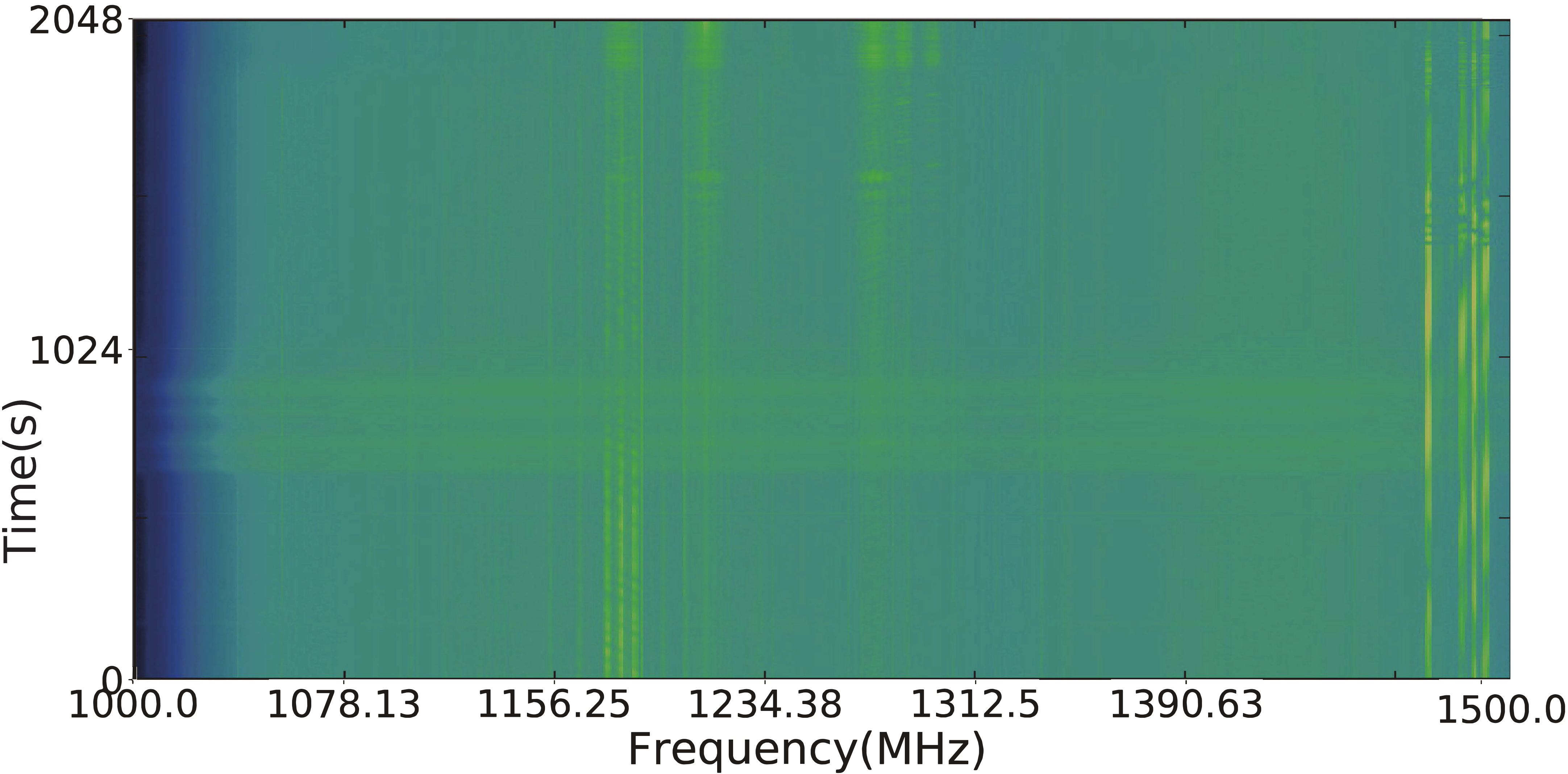}
		\caption{Data acquired by the FAST telescope shown in time-frequency plane. Bright stripes mark the distribution of radio frequency interference.}
		\label{Fig_1}
	\end{figure}

\begin{figure}
		\centering
		\includegraphics[width=0.9\linewidth]{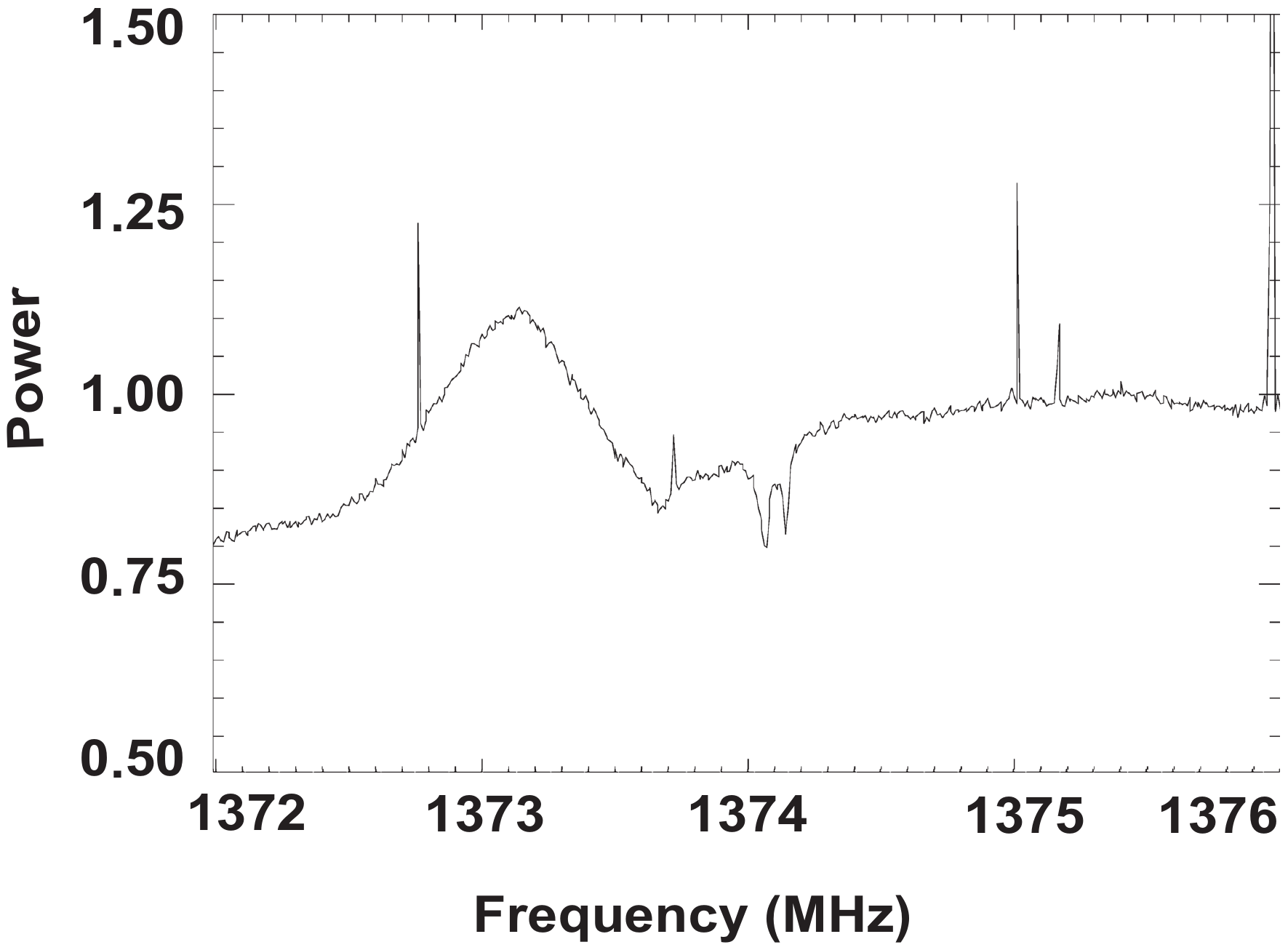}
		\caption{Example of sporadic baseline fluctuations in FAST's data. It can be seen that a wide bulge exists at $\sim 1373.2$ MHz. The amplitude of the bulge is obviously higher than background continuum, although still weaker than common RFI. Also, an extragalactic HI absorption line with double-horned structure can be seen at $\sim 1374.15$ MHz. The readings along y-axis are shown in normalised instrumental units without calibration.}
		\label{Fig_0}
	\end{figure}

	As can be seen in Fig.~\ref{Fig_1}, typical RFI can be much brighter than background noise or astronomical signals. Therefore, radio observations with a wide band coverage are easily affected by interference caused by human activities. Some weaker RFI may not be easily distinguished from celestial sources. Also, as occasionally seen in observed data of FAST, RFI may also occur in the form of fluctuations in baselines, appearing randomly at any time or frequency channel, as shown in Fig.~\ref{Fig_0}. It should be noted that with higher amplitudes (although much lower than typical RFI), such fluctuations should not be mistaken as standing waves \citep{1997PASA...14...37B} with quasi-stable periodic structures in the frequency domain \citep{2008A&A...479..903P}. Generally speaking, random fluctuations observed by FAST usually exhibit widths of the same order of magnitude as extragalactic HI lines, although they may show broad single-peaked structures, in contrast with HI line's double-horned ones. Therefore, although such random fluctuations may not be simply classified as RFI, we still need to mitigate the effects of such weak signals in radio astronomical data.

\begin{figure}
		\centering
		\includegraphics[width=\linewidth]{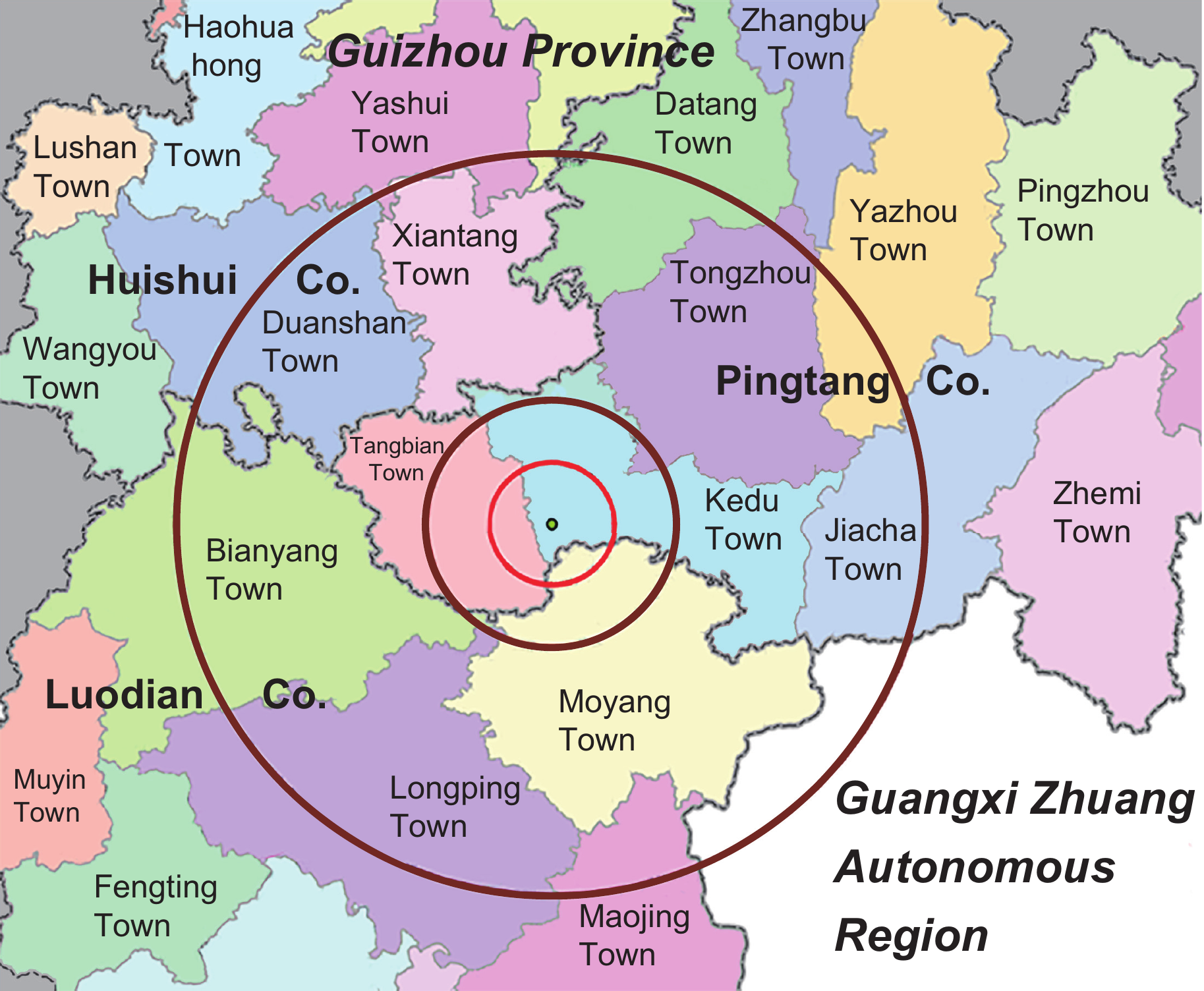}
		\caption{The radio quiet zone established by Guizhou Provincial People’s Government. With FAST site at its centre, the RQZ can be divided into a core area ($r \leq5$ km), an intermediate area ($r = 5-10$ km), as well as a remote area ($r = 10-30$ km), according to distance $r$ to the telescope. In the core area of this RQZ, it is forbidden to set up or use any radio stations, or to build and operate facilities that emit radio waves. In the intermediate area, it is prohibited to set up or use radio stations with working frequencies between $68-3000$ MHz, and effective radiation power higher than $100$ W. In the remote area, electromagnetic compatibility analysis should be conducted when setting up radio stations, or utilising facilities with working frequencies between $68-3000$ MHz, and effective radiation power higher than $100$ W. (Data source: \citet{Government2019}).}
		\label{Fig_zone}
	\end{figure}

	One of the most ideal and effective approaches to minimise the impact of RFI is to mitigate by establishing a Radio Quiet Zone (RQZ) surrounding a telescope to regulate RFI-emitting device operations in it \citep{AnTao2017}. For example, the Australian Communications and Media Authority (ACMA) has arranged a radio quiet zone in Western Australia \citep{7731554} for safety operations of radio astronomical instruments, including the low-frequency facility of the planned Square Kilometre Array (SKA). Similarly, With a government order named 'Regulations for Protection of Electromagnetic Wave Quiet Zone', Guizhou Province in China has also established an RQZ for the FAST telescope (see \citet{2013APS........14}, \citet{Government2019}, and Fig.~\ref{Fig_zone}), protecting its observing band from $70-3000$ MHz \citep{2011Nan}.
	
	Nevertheless, since emissions from sources like artificial satellites cannot be minimised by ground-based RQZs, and strong signals originating outside such areas can also be detected by radio telescopes, RFI detection still poses a challenge to radio observations, and the ability to detect RFI is an important issue for radio astronomical data reduction. The correctness and completeness of RFI flagging operations greatly affects the scientific output of each telescope. Moreover, the increase of human activities has boosted the complexity and occurrences of RFI, and complicated the task of RFI detections. Therefore, the aim of our study is to develop a suitable model to process the data acquired with FAST. Compared with other telescopes, such as Arecibo or Effelsberg, FAST has a superior sensitivity, thus rendering itself extremely vulnerable to RFI. As a result, FAST needs a more accurate RFI detection algorithm. Considering this telescope's various scientific goals, such as neutral hydrogen sky surveys as well as pulsar observations, the requirements of rapid adjustment and deployment should also be considered.
	
	Traditional RFI detection methods are mainly based on threshold algorithms \citep{2010MNRAS.405..155O, 2004AJ....128..933B}), as well as the physical characteristics of the RFI in the time-frequency domain (e.g. the linear detection, see \citet{Wolfaardt2016}). Related image-processing techniques are subsequently applied to improve the appearance on RFI detection edges. However, in actual applications, manual interventions are often required to refine or confirm the locations and ranges of interference, as well as to specify related parameters of the algorithm, thus greatly reducing the processing efficiencies. 
	
\begin{figure}
		\centering
		\includegraphics[width=\linewidth]{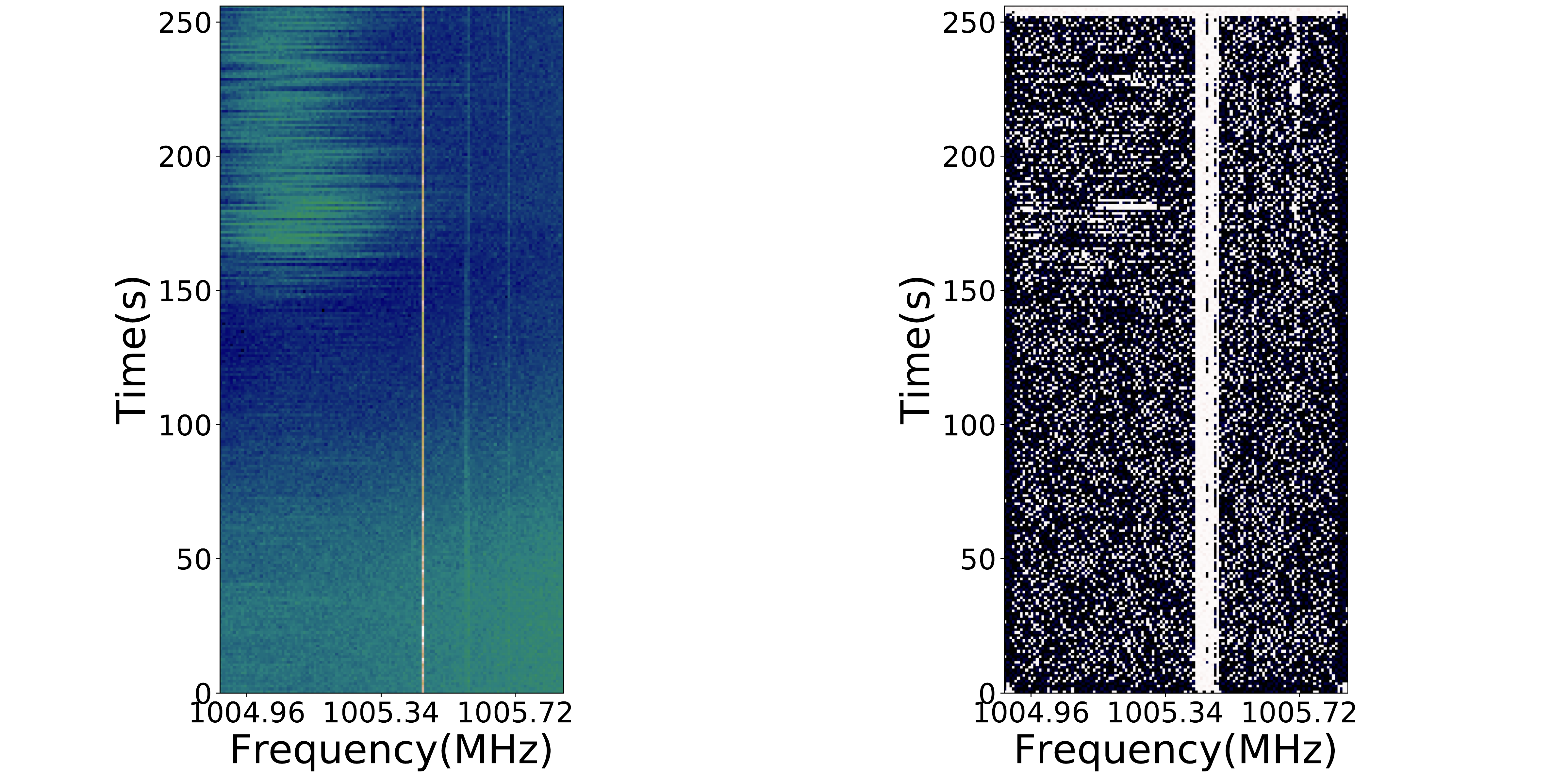}
		\caption{Detection result of U-Net (\citet{2017A&C....18...35A}) with observed data of FAST. The figure on the left shows the original data, and the figure on the right the detection result. It can be seen that the result looks like random noise. A large amount of RFI at the upper left corner and the bottom of the figure is not correctly detected, while too many false RFI signals are detected along the time axis (shown as vertical stripes).}
		\label{Fig_1_real_data_and_unet_result} 
	\end{figure}

	Currently, artificial intelligence (AI) technology represented by deep learning techniques has been used to detect RFI. \citet{2018AN....339..358B} has applied Recurrent Neural Network (RNN) to flag RFI, while \citet{2017A&C....18...35A} has performed similar tasks with Convolutional Neural Network (CNN). \citet{Czech2018} combined RNN and CNN to classify transient RFI sources. The application of AI technology has greatly reduced the workloads of astronomers, thus increasing the efficiency of data reduction. Yet, compared with traditional methods, the detection accuracies of existing deep learning algorithms still show very little advantage. Such methods usually have relatively low robustness, thus being unable to identify complicated RFI effectively. For example, it can be clearly seen that data shown in Fig.~\ref{Fig_1_real_data_and_unet_result} requires further processing. On the other hand, improving the accuracy blindly often leads to problems of overfitting. In addition, AI approaches could also be time consuming, and considerable efforts are required to prepare a large amount of training data to train the neural network. Thus, efficient and fast adaptation would be impractical for such algorithms.
	
	Inspired by these pioneering works, in this paper, we propose a new model which can improve the detection robustness without introducing artificial artefacts. Two types of residual learning \citep{2015arXiv151203385H, ARSALAN2019217} units have been adopted for down- and up-sampling processes in our CNN-based model, thus improving the accuracy without the requirements for large amounts of data. In addition, our model obviates the need for pre-treatments or further polishing, thereby increasing the efficiency and reliability of the process.
	
	Thus, the work presented in this paper can be summarized as:
	
	\begin{itemize}
		\item A network architecture named RFI-Net providing higher accuracy and minimal false-positives for RFI detection.
		\item Two types of residual units are utilised which can lead to equally or more accurate detections with less training data.
		\item A standalone method without the need for additional operations to achieve high efficiency and reliability is proposed.
	\end{itemize}
	
	Section~\ref{sec_2_RFI Recognition methods} provides a brief overview and analysis of related work. Section~\ref{sec_3_RFI-Net for recognition for FAST} presents the proposed RFI-Net with details of its features. Section~\ref{sec_4_Dataset and Experiment Framework} introduces the dataset and framework for all experiments carried out in this study. Section ~\ref{sec_5_Experiments and Results} presents the experiments and results. Our conclusions are given in Section~\ref{sec_6_Conclusion}.
	
	\section{RFI detection methods} 
	\label{sec_2_RFI Recognition methods}
	
	Generally speaking, RFI detection algorithms search for possible interference with specific signatures in observed data, and produce RFI masks marking positions of detected interference. Currently, traditional ways for RFI flagging include linear algorithms (i.e. SVD \citep{2010MNRAS.405..155O} and PCA \citep{Wolfaardt2016}), as well as threshold-based methods (SumThreshold \citep{2010MNRAS.405..155O} and CUMSUM \citep{2004AJ....128..933B}). Also, with the development of artificial intelligence in image recognition \citep{GOMEZRIOS2019315} and natural language processing \citep{EVANS2019353}, AI-related algorithms have been invoked by various branches in astronomy, from classifications of variable stars with enhanced performance in light-curve classification benchmarks \citep{2019MNRAS.482.5078A}, to pulsar candidate identifications \citep{2014Zhu}. Also, in the radio band, efforts have been made to apply CNN \citep{2017A&C....18...35A} and RNN \citep{Czech2018} for RFI detection.
	
	\subsection{Classical methods: SumThreshold as an example}
	\label{subsec_2.1_Recognition with classical methods}
	Since the signal strength of RFI is usually much stronger than that of typical astronomical signals, classical algorithms are based on physical characteristics of RFI. One of the notable approaches is SumThreshold, which is one of the most widely used algorithms \citep{2017A&C....18...35A}. Introduced by \citet{Offringa2010a}, the SumThreshold method has been proved to yield the highest accuracy among classical detection algorithms. SumThreshold can also be applied in combination with other algorithms, such as curvature fitting, to achieve better results \citep{Offringa2010a}.
	
	However, it is possible that the original SumThreshold method could mistake many 'good' samples as RFI, if no additional rules have been applied. Taking the dataset $[0,0,5,6,0,0]$ from \citet{2010MNRAS.405..155O} as an example, the data points with values of 5 and 6 contain strong interference. Since SumThreshold adopts a series of thresholds for average values of different-sized pixel combinations to identify RFI, we adopt a decreasing sequence $\chi_1=7, \chi_2=5, \chi_3=4, \cdots, \chi_6=1.8$ as thresholds of averaged values for $1, 2, 3, \cdots, 6$ pixels, respectively, as with \citet{2010MNRAS.405..155O}. With an averaged value larger than $6 \chi_6$, all six samples in the example dataset should be marked as RFI. In order to avoid such mislabelling, it is common practice to inspect pixel combinations with increasing sizes in the dataset. If a threshold $\chi_n$ for $n$-pixel combinations classifies a certain area as RFI, the readings of marked pixels should be replaced by $\chi_n$, leading the final average of our example to be $\frac{2\chi_6}{6}=0.6 < \chi_6$ \citet{2010MNRAS.405..155O}. However, the exact values of each threshold also need to be finely tuned, thus increasing the needs of manual interventions of this method. It is difficult to flag weak interference or unwanted baseline fluctuations with thresholding algorithms, since they may show flux levels similar to astronomical signals.
	
	\subsection{Detecting RFI with AI}
	\label{subsec_2.2_Recognition with AI}
	RFI flagging using CNN-based models has been studied in recent years (e.g., see \citet{2017A&C....18...35A}. In terms of accuracy, currently, the best variation of CNN model should be the U-Net model described in \citet{2017A&C....18...35A}. Here we present an overview of CNN, along with descriptions of the U-Net model.
	
	Computer vision adopts CNN as its main network scheme. The basic operation of CNN is convolution, in which the summed value of the product from the convolutional kernel is multiplying all sample values within an area covered by the kernel. Shallow convolutional operations extract textural information of images, while deep ones can integrate features obtained with shallow networks to get image semantics \citep{2015arXiv150504597R}. Thus, each layer of CNN performs convolution on one or more planes, extracting information from them, and applies pooling to reduce the volume of information. In this way, with multiple convolutions and pooling, specific information about certain areas of the image can be obtained. CNNs are suitable to make identifications in structural data, that is, data associated with spatially adjacent counterparts (e.g., images). 
	
	The U-Net model, originally proposed to meet the challenge as part of a workshop held prior to the IEEE International Symposium on Biomedical Imaging (ISBI) 2012 with notable results \citep{2015arXiv150504597R}, can be considered as CNNs with extended architectures, and have been adjusted for RFI detections \citep{2017A&C....18...35A}. It utilises down- and up-sampling operations to extract required information (such as RFI) from original data, enabling the network to actually 'learn' about the extracted characteristics. In this model, shallow-layered convolutions make identifications of fine features in data, e.g., RFI intersections, while deeper networks are used to splice such features into more abstract forms \citep{2017A&C....18...35A}. Meanwhile, features extracted by pooling operations during down-sampling are passed as copies to the right side after several steps of up-sampling, thus completing a U-shaped structure. With intensive tests, a structure with 3 layers and 64 feature graphs has achieved a good balance between flagging accuracy and computational cost. A structure like this can be seen in Figure 1 of \citet{2017A&C....18...35A}.

	\citet{2017A&C....18...35A} have also tested the U-Net model with data acquired by the Bleien Observatory \citep{BleienObservatoryData}. The visual inspections have been compared with results from U-Net, demonstrating the advantages of the latter technique with graphs as well as index scores. It has been proven that the U-Net shows an advantage over traditional RFI flagging algorithms. 
	
	\subsection{Analysis of previous methods}
	RFI with repetitive temporal or spectral behaviours, such as radar or radio beacon emissions, can be best identified using linear detection-based algorithms, although such methods are not suitable to detect stochastic RFI, including pulse-like signals, and RFI with frequency drifts\citep{2017A&C....18...35A}. In contrast, thresholding algorithms are more effective, if the observed background is relatively stable and the RFI is distributed discretely. Thresholding, with advantages of relatively fast execution speed, easy implementation, as well as high efficiency and robustness (with properly adjusted parameters), is best to flag strong RFI. And algorithms incorporating the SumThreshold method are especially popular in radio astronomy \citep{2017A&C....18...35A}. However, in the presence of weak interference or baseline fluctuations with fluxes comparable to celestial sources, or broad-band signals/extremely large amount of RFI (which means that most channels are RFI-contaminated), thresholding methods would become less effective, if not useless.
	
	On the other hand, methods invoking machine learning/deep learning techniques could greatly reduce manual involvements, thus enhancing the automation level of RFI detections. However, in terms of practical applications, these algorithms have not achieved satisfactory detection accuracy. As shown in Fig.~\ref{Fig_1_real_data_and_unet_result}, the U-Net model proposed by \citet{2017A&C....18...35A} has identified too many time domain structures as RFI; however, interference in the frequency domain, as well as point-like RFI, have been largely ignored. Furthermore, a lot of noise (also described as false positive) exists in the flagging results, especially at the edges of regions identified as RFI. Apparently, such performance needs to be further refined, which may decrease the efficiency of the algorithm as a result. Moreover, training neural networks like this often requires large amounts of pre-labelled data. While RFI flagging with raw data is a time-consuming and tedious task, considering the requirements for subsequent training, it would not be easy to make adjustments with existing approaches to improve their suitability for RFI detection in different observing configurations \citep{2019MNRAS.482.5078A}.
	
	\section{RFI-Net for RFI detections at FAST}
	\label{sec_3_RFI-Net for recognition for FAST}
	
\begin{figure}
		\centering
		\subfigure[]{\label{fig_3_simple_RFI-NET_structure}\includegraphics[width=\linewidth]{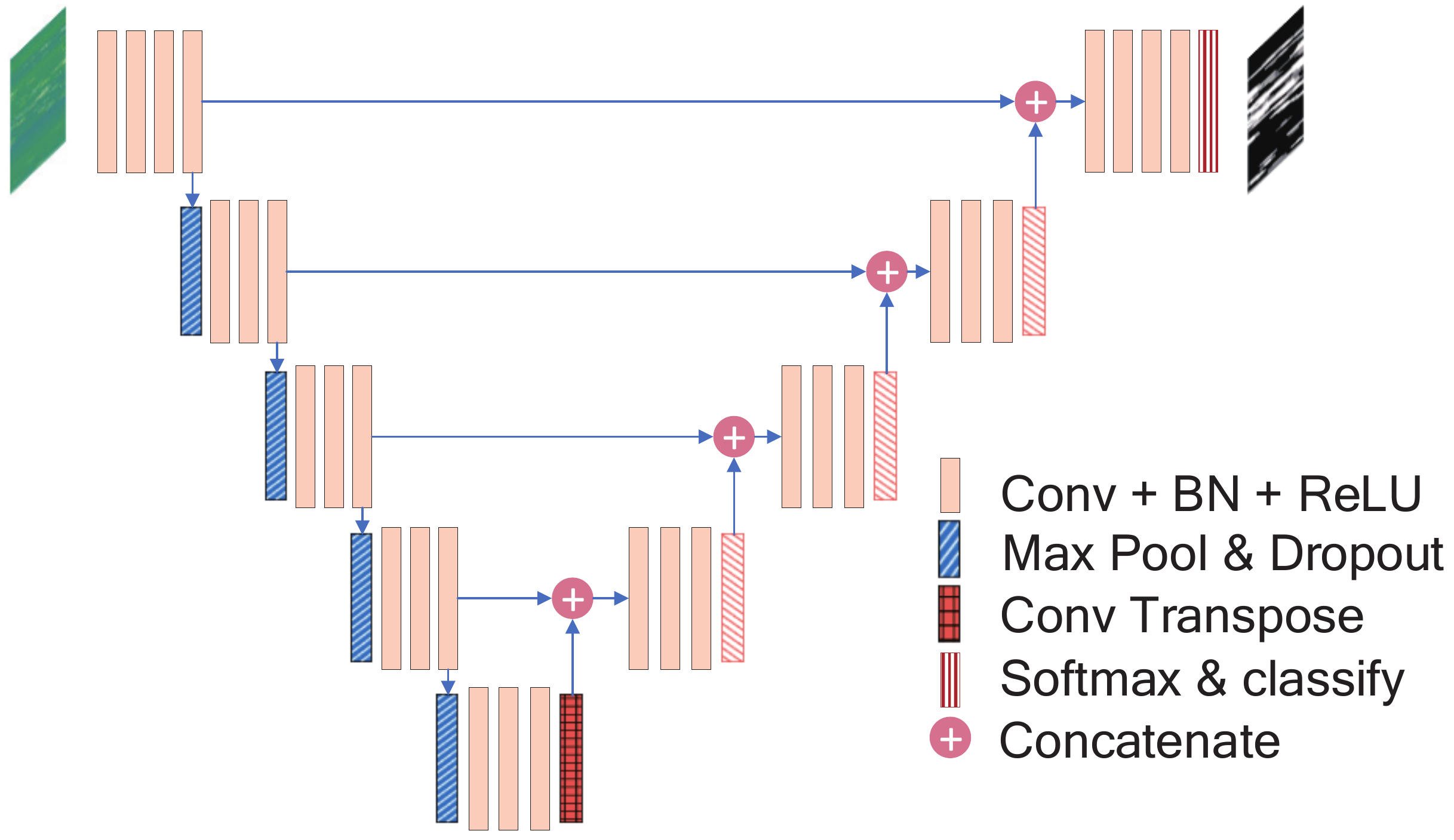}}
		\subfigure[]{\label{fig_3_RFI-NET_structure}\includegraphics[width=\linewidth]{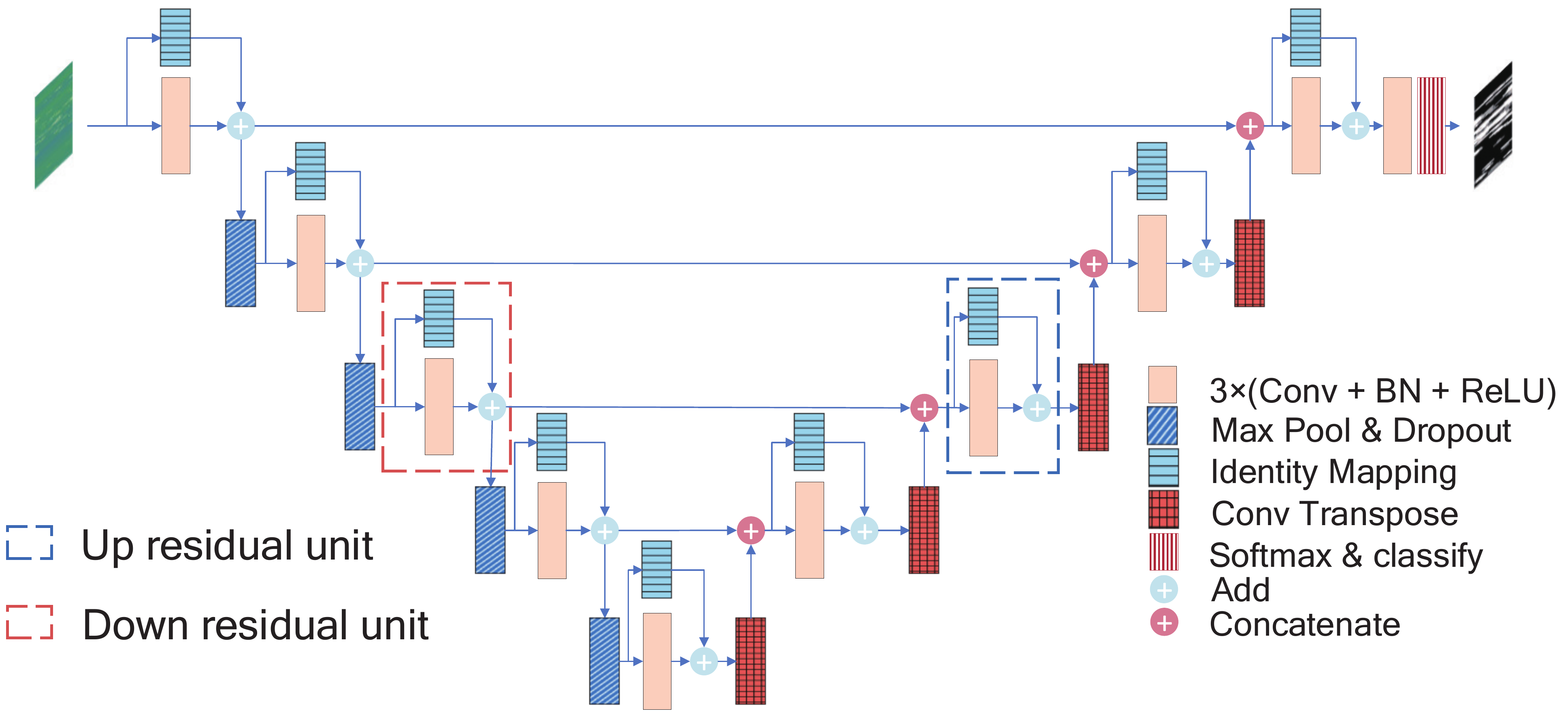}}
		\caption{Structure of RFI-Net. Here Conv denotes convolutional computation, BN the batch normalisation, ReLU the Rectified Linear Unit, which is a non-linear activation function \citep{2012:ICD:2999134.2999257}. Max Pool refers to a sample-based discretisation process for down-sampling of input representation, taking the maximum of the input, reducing its dimensionality, and making assumptions about features of the input data \citep{Boureau:2010:TAF:3104322.3104338}.  Dropout is used to prevent overfitting, as well as to reduce the time spent on training \citep{JMLR:v15:srivastava14a}. And identify Mapping makes it easier to train the network \citep{he2016identity}.}
		\label{Fig_3_net_structure} 
	\end{figure}

	To overcome the shortcomings of algorithms discussed so far, we propose a new model combining U-Net with a residual network. In this model, the corresponding layers are connected with short cuts, which are shown in Fig.~\ref{Fig_3_net_structure} as horizontal lines connecting the left and right sides. Two types of residual units have been designed and constructed. In addition, two other hyperparameters have been introduced to further enhance the performance\footnote{The codes related to this work are available at GitHub \url{https://github.com/DuFanXin/RFI-Net}}.
	
	\subsection{Architecture of RFI-Net}
	\label{subsec_3.1_Architecture of the network}
	We have constructed our model with two steps, the first one to add more layers to U-Net, as shown in Fig.~\ref{Fig_3_net_structure}(a). That is because, for deep learning algorithms, the depth of a certain neural network can be of great importance. Regardless of filter sizes or chosen widths, a deeper network  usually can achieve more optimal results, compared with shallower ones with roughly the same temporal complexity \citep{2014arXiv1412.1710H}, since an increase in depth can lead to more extracted information. Furthermore, compared with a shallower network, a deeper one can significantly enhance the capability to perform tasks that are more computationally demanding. 
	For the second step of our model construction, the residual units have been built into the network, as shown in Fig.~\ref{Fig_3_net_structure}(b). Identity mappings serve as short cuts to connect three convolutional layers. Two types of residual units have been designed in this work. The unit marked with orange dotted lines in Fig.~\ref{Fig_3_net_structure}(b) is used for down-sampling, thereby doubling the number of channels; whereas the unit indicated by blue dotted lines halves the number of channels for up-sampling. To support this structure, the network adopts batch normalisation to normalise the input data, and is equipped with a fast and stable optimiser. Detailed descriptions of the network structure are provided in Table~\ref{Tab_1_net_structure}.
	
	\begin{table*}
		\centering
		\caption{RFI-Net structure in detail. BN means Batch Normalisation for more stable data distributions, while transpose is sometimes called 'deconvolution' after Deconvolutional Networks, but it is in fact the transpose of the convolution. Most of the layers take ReLUs as activation functions to minimise the problem of gradient vanishing or explosion. In the last layer, SoftMax, the bracket means that if the network is in training process, it needs SoftMax only. Otherwise, the ArgMax should also be processed to get RFI flagging result.}
		\label{Tab_1_net_structure}
		\resizebox{\linewidth}{!}{
			\begin{tabular}{cccccc}
				\hline
				layer & \multicolumn{2}{c}{operation} & input size & filter size & output size \\
				\hline
				conv 1 & \multicolumn{2}{c}{Conv+BN+ReLU} & $256 \times 128 \times 1$ & $3 \times 3 \times 1\times32$ & $256 \times 128 \times 32$ \\
				conv 2 & \multicolumn{2}{c}{Conv+BN+ReLU} & $256 \times 128 \times 32$ & $3 \times 3 \times 1\times32$ & $256 \times 128 \times 32$ \\
				down residual unit 1 & Conv+BN+ReLU & \multirow{3}[0]{*}{Conv+BN} & $256 \times 128 \times 32$ &  & $256 \times 128 \times 64$ \\
				& Conv+BN+ReLU &       &       &       &  \\
				& Conv+BN &       &       &       &  \\
				max pool 1 & \multicolumn{2}{c}{Max pooling+Dropout} & $256 \times 128 \times 64$ & $1 \times 2 \times 2\times1$ & $128 \times 64 \times 64$ \\
				down residual unit 2 & Conv+BN+ReLU & \multirow{3}[0]{*}{Conv+BN} & $128 \times 64 \times 64$ & - & $128 \times 64 \times 128$ \\
				& Conv+BN+ReLU &       &       &       &  \\
				& Conv+BN &       &       &       &  \\
				max pool 2 & \multicolumn{2}{c}{Max pooling+Dropout} & $128 \times 64 \times 128$ & $1 \times 2 \times 2\times1$ & $64 \times 32 \times 128$ \\
				down residual unit 3 & Conv+BN+ReLU & \multirow{3}[0]{*}{Conv+BN} & $64 \times 32 \times 128$ & - & $64 \times 32 \times 256$ \\
				& Conv+BN+ReLU &       &       &       &  \\
				& Conv+BN &       &       &       &  \\
				max pool 3 & \multicolumn{2}{c}{Max pooling+Dropout} & $64 \times 32 \times 256$ & $1 \times 2 \times 2\times1$ & $32 \times 16 \times 256$ \\
				down residual unit 4 & Conv+BN+ReLU & \multirow{3}[0]{*}{Conv+BN} & $32 \times 16 \times 256$ & - & $32 \times 16 \times 512$ \\
				& Conv+BN+ReLU &       &       &       &  \\
				& Conv+BN &       &       &       &  \\
				max pool 4 & \multicolumn{2}{c}{Max pooling+Dropout} & $32 \times 16 \times 512$ & $1 \times 2 \times 2\times1$ & $16 \times 8 \times 512$ \\
				down residual unit 5 & Conv+BN+ReLU & \multirow{3}[0]{*}{Conv+BN} & $16 \times 8 \times 512$ & - & $16 \times 8 \times 1024$ \\
				& Conv+BN+ReLU &       &       &       &  \\
				& Conv+BN &       &       &       &  \\
				up sample 1 & \multicolumn{2}{c}{Transpose+BN+ReLU} & $16 \times 8 \times 102$4 & $1 \times 2 \times 2\times1$ & $32 \times 16 \times 512$ \\
				concat 1 & \multicolumn{2}{c}{Concatenate} & $2\times(32 \times 16 \times 512$) & -     & $32 \times 16 \times 102$4 \\
				up residual unit 1 & Conv+BN+ReLU & \multirow{3}[0]{*}{Conv+BN} & $32 \times 16 \times 102$4 & - & $32 \times 16 \times 512$ \\
				& Conv+BN+ReLU &       &       &       &  \\
				& Conv+BN &       &       &       &  \\
				up sample 2 & \multicolumn{2}{c}{Transpose+BN+ReLU} & $32 \times 16 \times 512$ & $1 \times 2 \times 2\times1$ & $32 \times 16 \times 256$ \\
				concat 2 & \multicolumn{2}{c}{Concatenate} & $2\times(64 \times 32 \times 256$) & -     & $64 \times 32 \times 512$ \\
				up residual unit 2 & Conv+BN+ReLU & \multirow{3}[0]{*}{Conv+BN} & $64 \times 32 \times 512$ & - & $64 \times 32 \times 256$ \\
				& Conv+BN+ReLU &       &       &       &  \\
				& Conv+BN &       &       &       &  \\
				up sample 3 & \multicolumn{2}{c}{Transpose+BN+ReLU} & $64 \times 32 \times 256$ & $1 \times 2 \times 2\times1$ & $128 \times 64 \times 128$ \\
				concat 3 & \multicolumn{2}{c}{Concatenate} & $2\times(128 \times 64 \times 128$) & -     & $128 \times 64 \times 256$ \\
				up residual unit 3 & Conv+BN+ReLU & \multirow{3}[0]{*}{Conv+BN} & $128 \times 64 \times 256$ & - & $128 \times 64 \times 128$ \\
				& Conv+BN+ReLU &       &       &       &  \\
				& Conv+BN &       &       &       &  \\
				up sample 4 & \multicolumn{2}{c}{Transpose+BN+ReLU} & $128 \times 64 \times 128$ & $1 \times 2 \times 2\times1$ & $256 \times 128 \times 64$ \\
				concat 4 & \multicolumn{2}{c}{Concatenate} & $2\times(256 \times 128 \times 64$) & -     & $256 \times 128 \times 128$ \\
				up residual unit 4 & Conv+BN+ReLU & \multirow{3}[0]{*}{Conv+BN} & $256 \times 128 \times 128$ & - & $256 \times 128 \times 64$ \\
				& Conv+BN+ReLU &       &       &       &  \\
				& Conv+BN &       &       &       &  \\
				conv 2 & \multicolumn{2}{c}{Conv+BN} & $256 \times 128 \times 64$ & $3 \times 3 \times 64 \times 2$ & $256 \times 128 \times 2$ \\
				softmax & \multicolumn{2}{c}{SoftMax(+ArgMax)} & $256 \times 128 \times 2$ & -     & $256 \times 128 \times 1$ \\
				\hline
			\end{tabular}
		}
	\end{table*}
	
	\subsection{Residuals Unit}
	\label{subsec_3.2_Residuals Unit}
	The residual units have been added to prevent network degeneration. When a deep network no longer diverges despite implementations of various optimisations, it will begin to degrade: with an increasing network depth, the detection accuracy would finally meet a 'bottleneck', with rapidly growing training errors afterwards. It should be noted that such errors introduced by network degradation are due to more biased calculations caused by a larger number of layers, rather than overfitting \citep{2015arXiv151203385H}. 
	
	\begin{table}
		\centering
		\caption{Detailed hyperparameters of the residual unit for down-sampling. Here $H$ represents the height of the data, $W$ the data width, and $C$ the number of channels. The input data of 'short cut' and ‘conv 1’ layers are identical. The layer 'add' has two sources of inputs: layers 'conv 1' and 'short cut'.}
		\label{Tab_2_residual_unit_down}
		\resizebox{\linewidth}{!}{
			\begin{tabular}{cccccc}
				\hline
				layer & operation & input size & filter size & output size \\
				\hline
				conv 1 & Conv+BN+ReLU & $H \times W \times C$ & $3 \times 3 \times C \times 2C$ & $H \times W \times 2C$ \\
				conv 2 & Conv+BN+ReLU & $H \times W \times 2C$ & $3 \times 3 \times 2C \times 2C$ & $H \times W \times 2C$ \\
				conv 3 & Conv+BN & $H \times W \times 2C$ & $3 \times 3 \times 2C \times 2C$ & $H \times W \times 2C$ \\
				short cut & Conv+BN & $H \times W \times C$(same as conv 1 input) & $1 \times 1 \times C \times 2C$ & $H \times W \times 2C$ \\
				add & Add+BN+ReLU & $H \times W \times 2C(\text{from conv 1 result})$ & - & $H \times W \times 2C$ \\
				&  &  $+ H \times W \times 2C(\text{from short cut result}) $&       &  \\
				\hline
			\end{tabular}%
		}
	\end{table}%
	
	\begin{table}
		\centering
		\caption{Detailed hyperparameter of residual unit for up-sampling. Here $H$ represents the height of the data, $W$ the width data, and $C$ the number of channels. This unit halves the channel number of the data.}
		\label{Tab_3_residual_unit_up}
		\resizebox{\linewidth}{!}{
			\begin{tabular}{cccccc}
				\hline
				layer & operation & input size & filter size & output size \\
				\hline
				conv 1 & Conv+BN+ReLU & $H \times W \times C$ & $3 \times 3 \times C \times C/2$ & $H \times W \times C/2$ \\
				conv 2 & Conv+BN+ReLU & $H \times W \times C/2$ & $3 \times 3 \times C/2 \times C/2$ & $H \times W \times C/2$ \\
				conv 3 & Conv+BN & $H \times W \times C/2$ & $3 \times 3 \times C/2 \times C/2$ & $H \times W \times C/2$ \\
				short cut & Conv+BN & $H \times W \times C$(same as conv 1 input) & $1 \times 1 \times C \times C/2$ & $H \times W \times C/2$ \\
				add & Add+BN+ReLU & $H \times W \times C/2(\text{from conv 1 result})$ & - & $H \times W \times C/2$ \\
				&  &  $+ H \times W \times C/2(\text{from short cut result})$ &       &  \\
				\hline
			\end{tabular}%
		}
	\end{table}%
	
	Therefore, inspired by the residual network \citep{2015arXiv151203385H}, we introduce short cuts to our model, which are represented by the dotted boxes in Fig.~\ref{Fig_3_net_structure}
	\begin{equation}
	\label{Eq_3_con}
	\centering
	z = F(x)
	\end{equation}
	where $x$ and $z$ denote the input and output, respectively. Identity mapping in most residual networks can be represented as follows
	
	\begin{equation}
	\label{Eq_3_identity_map}
	\centering
	z = F(x) + x
	\end{equation}
	
	Equation \ref{Eq_3_identity_map} expresses that the input should be directly added to the result of the convolutional operations. However, since identity mapping does not have the flexibility to resize, especially with respect to channel numbers, RFI-Net chooses to perform convolutional operations with a kernel size of $1\times1$. With the assistance of batch normalisation, the short cut can be expressed as
	
	\begin{equation}
	\label{Eq_3_real_short_cut}
	\centering
	z = F(x) + H(x)
	\end{equation}
	where $H(x)$ denotes the combination of $1\times1$ convolution, as well as batch normalisation. Our network design adopts two units corresponding to the processes of down-and up-sampling to make adjustment of channel numbers possible. Details of the unit hyperparameters are presented in Tables~\ref{Tab_2_residual_unit_down} and \ref{Tab_3_residual_unit_up}.
	
	By connecting three layers as a unit, the short cuts can stabilise the update process, and slow down the gradient disappearance resulting from the inability of the model's middle layers to update the parameters effectively. Moreover, it is expected that models with short cuts should not only see improvements of their performance, but also achieve a faster convergence speed \citep{Drozdzal2016TheIO}. In addition, the long connecting structures used in the U-Net model could give rise to a slower learning rate with unstable parameter updates \citep{Drozdzal2016TheIO}. In contrast, a network equipped with short cuts makes a larger initial learning rate possible, thus accomplishing a faster convergence.
	
	\subsection{Additional hyperparameters}
	\label{subsec_3.3_Additional hyperparameters}
	Batch Normalisation, which is generally referred to as batch-norm, is achieved by using an algorithm to normalise batches of data with adjustments of averaged data to 0, and the variance to 1. Let $m$ be the size of input sample $X=(x_1,\cdots,x_m)$, the  normalised data should be calculated as:
	
	\begin{equation}
	\label{Eq_3}
	\centering
	\mu=\frac{1}{m}\sum_{i=1}^{m}x_i
	\end{equation}
	
	\begin{equation}
	\label{Eq_4}
	\centering
	\sigma^2=\frac{1}{m}\sum_{i=1}^mx_i^2
	\longrightarrow
	\sigma^2=\frac{1}{m}\sum_{i=1}^m(x_i-\mu)^2
	\end{equation}
	
	\begin{equation}
	\label{Eq_5}
	\centering
	\overline{x_i}=\frac{x_i-\mu}{\sqrt{\mu^2+\varepsilon}}
	\longrightarrow
	\overline{x_i}=\frac{x_i-\mu}{\sqrt{\sigma^2+\varepsilon}}
	\end{equation}
	
	Equation \ref{Eq_3} calculates the mean value of the sample, with Equation \ref{Eq_4} providing the sample variance. And Equation \ref{Eq_5} subtracts the mean value from each data point in the sample, dividing the results with the variance, thus completing the normalisation process. The bias $\varepsilon$ is introduced in Equation \ref{Eq_5} to prevent the occurrence of zero in the divisor. Without batch normalisation, The neural network needs to constantly update the parameters with continuous back-propagation to obtain results from each layer. Since the distribution of such results could change continually \citep{2015arXiv150203167I}, time-to-time re-adjustments of parameters from the previous layer to accommodate new distributions are required, thus preventing the model achieving a faster learning rate during training. The application of batch normalisation ensures that the resulting distribution of each layer can be nearly constant with time, thus avoiding the need to re-adjust the previous layer. However, since each layer is normalised, the absolute values of the results are small, with results from the front layers differing insignificantly from those of the back layers, which may lead the network model to 'learn' features ineffectively  \citep{2015arXiv150203167I}. Thus, a controllable adjustment method to appropriately scale the normalised results is proposed with equation \ref{Eq_6}.
	
	\begin{equation}
	\label{Eq_6}
	\centering
	y_i=\gamma\overline{x_i}+\beta
	\end{equation}
	
	In equation \ref{Eq_6}, $\gamma$ has the task of scaling sizes of the normalised results, whereas $\beta$ shifts the result. Both of these parameters can make self-learning possible. Scaling is used to restore the results to its original value at least \citep{2015arXiv150203167I}. This approach based on batch normalisation enables the model's learning rate to be accelerated without the need to impose strict parameter initialisation, thus promoting faster convergence.
	
	The optimiser strategy we adopted here to set the mini-batch size in the hyperparameter is called Small Batch Gradient Descent, which is used in combination with the Adam algorithm \citep{Kingma2014} as the computing strategy for gradient descent. This strategy can speed up the gradient descent, and minimise oscillation of the gradient.
	
	The small Batch Gradient Descent is based on the stochastic gradient descent method \citep{2016arXiv160904747R}, which adopts a sample size between that of the batch and the random gradient drop. This method selects samples from data to calculate the gradient, and then performs the gradient drop. This approach not only enables the gradient to decrease at a faster rate, but also ensures that the gradient remains relatively stable during gradient descent.
	
	The Adam algorithm \citep{Kingma2014}, which combines Momentum \citep{2017arXiv170303633L} with RMSprop \citep{2016arXiv160904747R}, uses both first- and second-order moments, and performs a similar standardised operation on the gradient. The main advantage of the Adam algorithm is that, after offset correction, the learning rate of each iterative cycle can fall within a certain range, thus stabilising the parameters. This algorithm can also calculate different adaptive learning rates for different parameters. Experiences have shown that the Adam can perform well in practice, and is more computationally efficient compared with other adaptive learning algorithms \citep{2016arXiv160904747R}.

	\section{Basic Framework of Data and Experiments}
	\label{sec_4_Dataset and Experiment Framework}
	
	We now describe our experiments. Section~\ref{subsec_4.1_Dataset for experiment} discusses the data used for training and evaluation, and the experimental framework is described in Section~\ref{subsec_4.2_Experiment Framework}, which compares the models employed in the experiments, discusses the software and hardware used in the experiments, and presents an evaluation of the model performance.
	
	\subsection{Data used in the experiments}
	\label{subsec_4.1_Dataset for experiment}
	We firstly adopted an astronomical simulator to simulate data captured by a radio telescope. The FAST telescope can be used to conduct observations of neutral hydrogen (HI), and the 21-cm line from the hyperfine transition of neutral hydrogen emitted at a rest frequency of $\sim 1420.4$ MHz. Many software packages are available for HI simulation and data processing. The simulator we used for this study is HIDE (the HI Data Emulator), which was also used in the studies of \citet{2017A&C....18...35A} to simulate the training dataset.
	
	The simulated data are comprised of astronomical data and RFI. Since HIDE can produce both 'pure' RFI and simulated astronomical data (that already contain RFI), as shown in the top left panel of Fig.~\ref{Fig_4_result_in_visual}, it is possible to label all the interference precisely as fundamental references (that is, the ground\_truth) for our experiments. By comparing RFI detected by various algorithms with the ground\_truth, the accuracy of each method can be evaluated. The simulated RFI is displayed in the top centre panel of Fig.~\ref{Fig_4_result_in_visual}, with a corresponding RFI mask shown in the top right.
	
	Experiments have also been conducted with manually-labelled observed datasets captured by FAST in September, 2018, as well as open access data \citep{BleienObservatoryData} acquired by the Bleien Observatory on March 21st, 2016 \citep{2017A&C....18....8A}.
	
	\subsection{Experimental framework}
	\label{subsec_4.2_Experiment Framework}
	For the purpose of comparison, our dataset has been processed by several existing RFI flagging methods, including U-Net, KNN (K-Nearest Neighbour \citep{10.1007/978-3-540-39964-3_62}, one of the classification algorithms used in machine learning), as well as SumThreshold. Similar to \citet{2017A&C....18...35A}, we have found that U-Net with three layers and 64 characters can achieve a good balance between accuracy and speed, after several cycles of tests. Thus, in this study, we have adopted the same structure and hyperparameters for RFI-Net as \citet{2017A&C....18...35A}. In addition, this study applied Scikit-learn \citep{2011scikitlearn11}, the machine-learning library for Python, to assist with the implementation of the KNN algorithm.
	
	Specifically, the following experiments have been conducted: 1) We compared the results obtained from RFI-Net for simulated data with those from previous methods; 2) a detailed analysis of algorithm accuracies was carried out; 3) the results obtained from RFI-Net for observed data were studied to determine whether additional operations were required; 4) We use RFI-Net to process less training data (i.e. 25\%, 50\%, and 75\% of the complete data) to validate its ability to achieve comparable performance on smaller datasets; 5) the ability of RFI-Net to overcome the overfitting problem was demonstrated; 6) the training speed was investigated.

	All these methods have been tested under Ubuntu 16.04 on a Dell PowerEdge T630 with 32 GB RAM. Those deep learning models have been executed on a NVIDIA Tesla K40c with 12 GB RAM.
	
	The indicators we adopted for experimental evaluations are precision, recall, as well as the F1 score see \citet{2017A&C....18...35A} and \citet{davis2006relationship}:
	
	Precision, which is the fraction of genuine RFI among all flagged instances, can be considered as the accuracy metric of the algorithm
	\begin{equation}
	\label{Eq_precision}
	\centering
	Precision = \frac{\text{True Positive}}{\text{True Positive} + \text{False Positive}}
	\end{equation}

	Recall indicates the fraction of RFI that have been identified among all RFI, showing the comprehensiveness of the detection
	\begin{equation}
	\label{Eq_recall}
	\centering
	Recall = \frac{\text{True Positive}}{\text{True Positive} + \text{False Negative}}
	\end{equation}
	
	F1, the reciprocal mean of precision and recall, can be considered as the overall model performance
	\begin{equation}
	\label{Eq_f1}
	\centering
	F1 = \frac{2 \times \text{Precision} \times \text{Recall}}{\text{Precision} + \text{Recall}}
	\end{equation}
	
	It should be noted that, since our observed data were labelled manually, with no guarantee of completeness of RFI flags, the related performances cannot be evaluated exactly with indicators listed above. They rely on visual inspections, and can only provide a general approximation of algorithm characteristics.
	
	\section{Results of Experiments}
	\label{sec_5_Experiments and Results}

	In this section, we compare the visual results obtained by our model with those of previous methods in Section~\ref{subsec_5.1_Comparison of the RFI-Net with previous methods}. This is followed by Section~\ref{subsec_5.2_Impact on accuracy}, which proves the advantages of using detailed indicators. The test results with real observing data are presented in Section \ref{subsec_5.3_Dependence on extra processing} and the experimental results obtained with smaller training dataset are listed in Section~\ref{subsec_5.4_Accuracy on less training data}. Section~\ref{subsec_5.5_Impact on overfitting} demonstrates the impacts of the overfitting problem, while the ability of RFI-Net to accelerate the training process is presented in Section~\ref{subsec_5.5_RFI-Net on training speed}.  
	
	\subsection{Comparison of RFI-Net with previous methods}
	\label{subsec_5.1_Comparison of the RFI-Net with previous methods}
	
	\begin{figure}
		\centering
		\includegraphics[width=\linewidth]{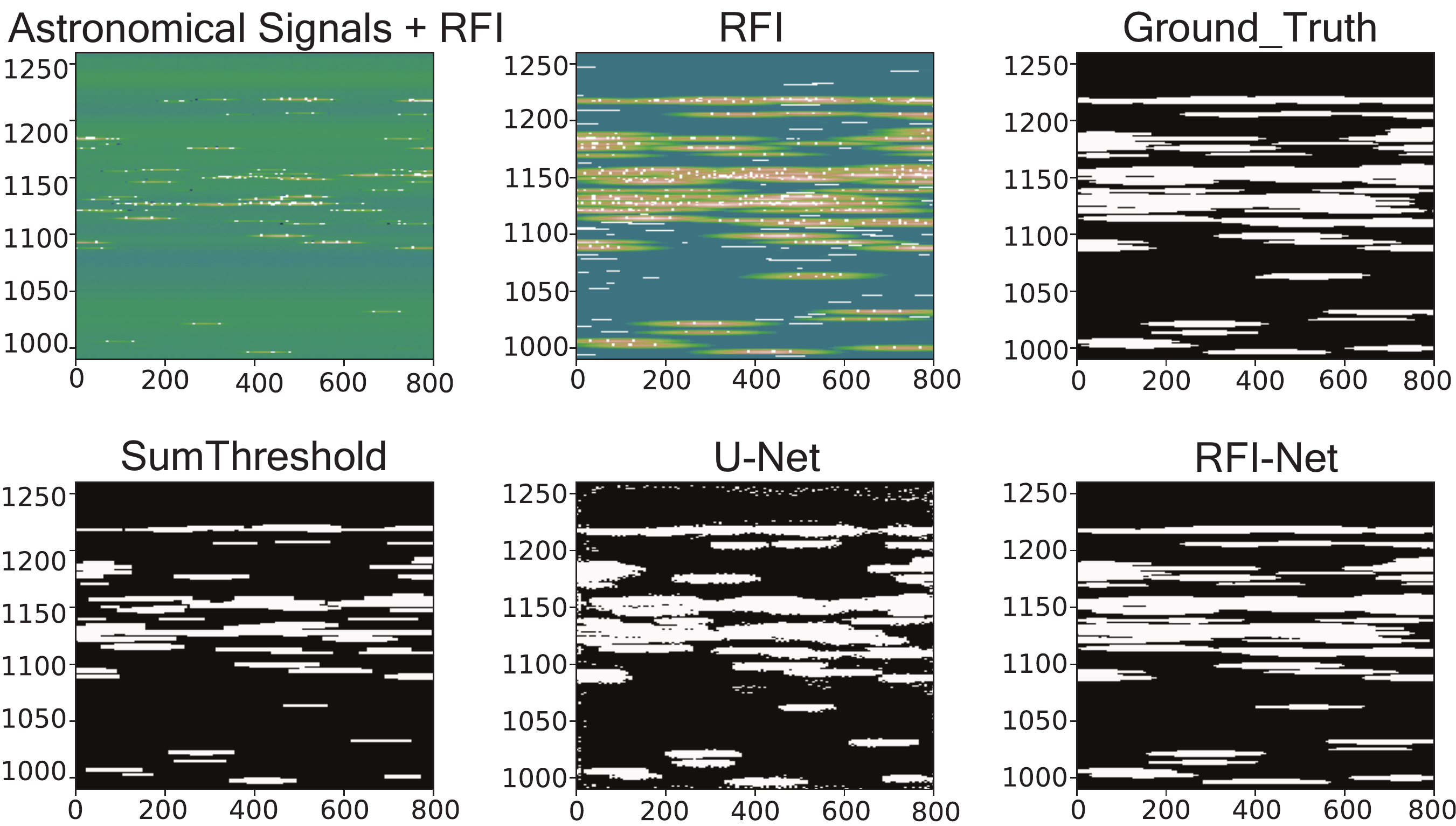}
		\caption{Simulated data with identified RFI masks obtained using various algorithms. From left to right, top images present the simulated astronomical data which contain signals and RFI, RFI, as well as corresponding RFI mask served as ground\_truth, in which white denotes RFI, and black for 'good' data, as generated by HIDE \citep{2017A&C....18....8A}. Bottom images show the results from different methods. It can be seen that the one obtained with our method resembles most closely to the ground\_truth. The result of U-Net is not so clean, while that of SumThreshold fails to detect some of the RFI.}
		\label{Fig_4_result_in_visual}
	\end{figure}
	
	The results of our RFI-Net model, the U-Net model, as well as SumThreshold are shown in Fig.~\ref{Fig_4_result_in_visual}, with identified RFI of each method listed as RFI masks. It can be clearly seen that the result of the model proposed in this paper shows the greatest resemblance  to the ground\_truth of our simulated data. The RFI-Net model correctly labelled most RFI, with only a small amount of data misidentified. A few flaws in our result only show up in certain details (such as at the edge portion of some RFI). By comparison, although the detection result of the SumThreshold algorithm seems to be relatively clean, with fewer cases of false RFI detections, some of the RFI is not correctly detected when compared with the ground\_truth, resulting in a lower the recall rate. And for the U-Net model, too many false positives have been identified, despite a number of RFI remaining undetected. In addition, U-Net erroneously labels a lot of 'good' pixels as RFI, which means its detection result is less 'clean' than our model.
	
	\subsection{Comparison of accuracy}
	\label{subsec_5.2_Impact on accuracy}
	
	As illustrated in section \ref{subsec_4.2_Experiment Framework}, the performances of various algorithms can be measured with scores based on parameter precision, recall, as well as F1 score. Table~\ref{Tab_4_performance_of_methods} lists the results of our RFI-Net model, the U-Net model, KNN, as well as SumThreshold with simulated data, indicating that RFI-Net has outperformed two other methods, with abilities to detect RFI more accurately and comprehensively. Although the result obtained with SumThreshold is as accurate as ours in terms of precision, it failed to detect RFI consistently. And although the performance of the CNN-based U-Net method shows a more balanced performance among all parameters, the detection quality of its results cannot be compared with those of RFI-Net. KNN, on the other hand, shows the worst performance in terms of all three parameters. In general, these results show the application of AI for RFI detection tasks to be promising, and our model has a better performance than previous algorithms.
	
	\begin{table}
		\centering
		\caption{Detailed scores of algorithm performances obtained with simulated dataset. The precision represents accuracy, recall shows detecting comprehensiveness, while the F1 score is the harmonic mean of these two indicators. The highest scores for each indicator are shown in bold font.}
		\label{Tab_4_performance_of_methods}
		\begin{tabular*}{\linewidth}{@{\extracolsep{\fill}}cccc}
			\hline  
			& Precision & Recall & F1 Score\\
			\hline  
			RFI-Net & \textbf{97.93} & \textbf{95.43} & \textbf{96.67}\\
			SumThreshold & 97.20 & 58.64 & 73.15\\
			KNN & 76.83 & 68.77 & 72.57\\
			U-Net & 80.87 & 81.30 & 81.04\\
			\hline
		\end{tabular*}
	\end{table}
	
	In Table~\ref{Tab_5_performance_of_methods}, RFI-Net demonstrates the ability to detect RFI more accurately in real data. Although here the RFI is more complex than in the simulated data, RFI-Net has still done the best job. 
	
	\begin{table}
		\centering
		\caption{Detailed scores of performances tested with FAST's observation. The best scores of each indicator are shown in bold font.}
		\label{Tab_5_performance_of_methods}
		\begin{tabular*}{\linewidth}{@{\extracolsep{\fill}}cccc}
			\hline  
			& Precision & Recall & F1 Score\\
			\hline  
			RFI-Net & \textbf{97.34} & 89.12 & \textbf{93.05}\\
			SumThreshold & 91.20 & 59.63 & 72.11\\
			KNN & 87.69 & \textbf{94.05} & 90.75\\
			U-Net & 95.11 & 88.19 & 91.52\\
			\hline
		\end{tabular*}
	\end{table}
	
	\begin{table}
		\centering
		\caption{Detailed performances obtained using Bleien's observations. The highest scores are shown in bold font.}
		\label{Tab_6_performance_of_methods}
		\begin{tabular*}{\linewidth}{@{\extracolsep{\fill}}cccc}
			\hline  
			& Precision & Recall & F1 Score\\
			\hline  
			RFI-Net & \textbf{95.61} & 91.07 & \textbf{93.28}\\   
			SumThreshold & 87.26 & 58.64 & 70.14\\
			KNN & 70.58 & \textbf{97.60} & 81.36\\
			U-Net & 94.08 & 89.14 & 91.54\\
			\hline
		\end{tabular*}
	\end{table}

	As shown in Table~\ref{Tab_6_performance_of_methods}, RFI-Net has achieved the highest precision and F1 score when tested with data from Bleien Observatory. KNN scored the highest accuracy, $97.60$, along with a much lower comprehensiveness. In contrast, SumThreshold failed to achieve good results, and ranked the last with worst recall and F1 score. U-Net has performed well, with all three indicators scoring about $90$. Nevertheless, these scores imply that there is still room to further improve the performance of the RFI-Net model.
	
	\subsection{Dependence on extra processing}
	\label{subsec_5.3_Dependence on extra processing}
\begin{figure}
		\centering
		\includegraphics[width=\linewidth]{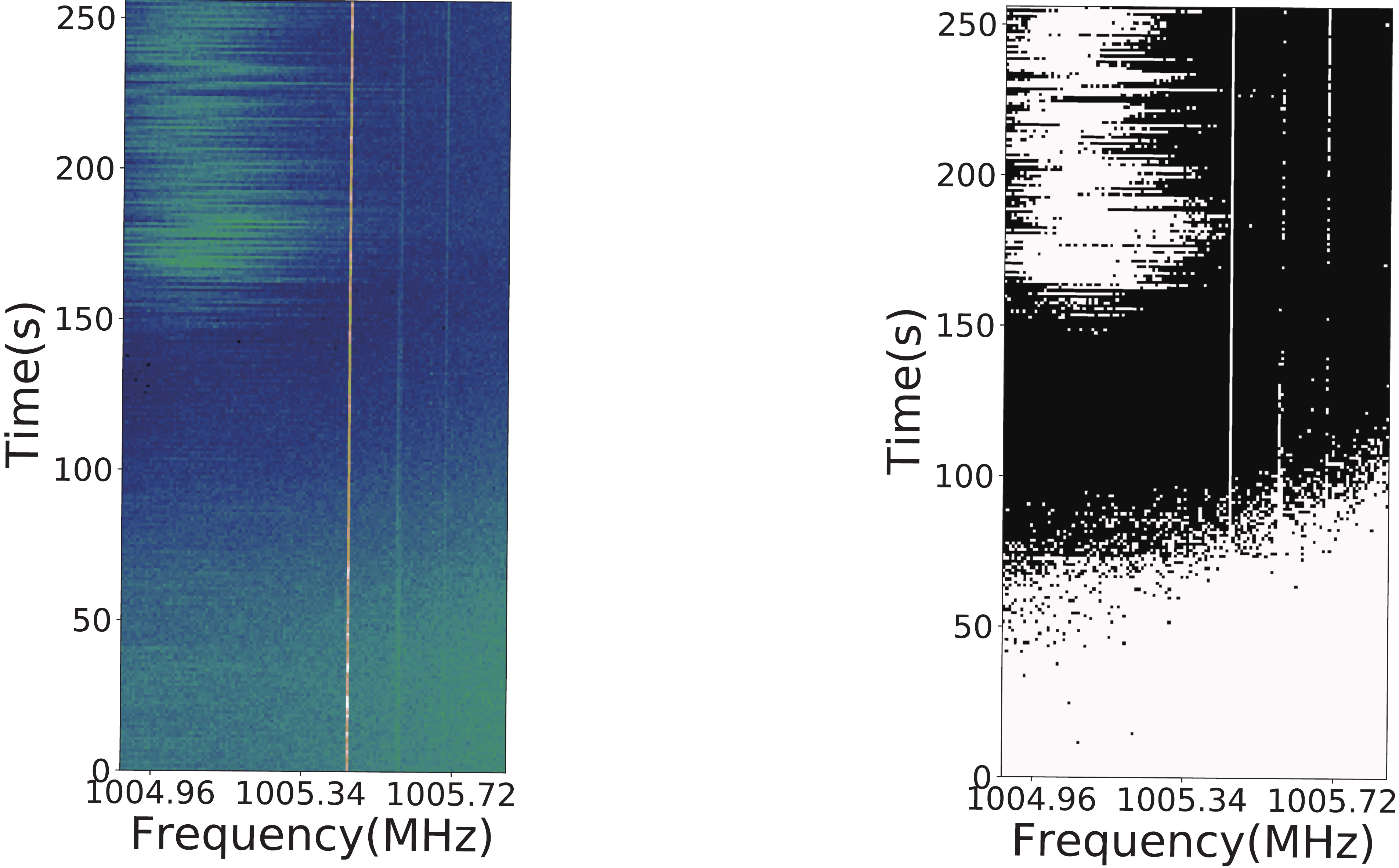}
		\caption{Left: Result obtained with FAST observations, which is the same as Fig.~\ref{Fig_1_real_data_and_unet_result}. Most RFI in frequency and time domains, as well as point-like disturbances can be detected. However, some of the indistinct RFI has been missed. Right: RFI flagging mask of FAST observations. Conspicuous RFI has been successfully identified, whereas some indistinct RFI remains undetected.}
		\label{Fig_4_real_data_rfi-net} 
	\end{figure}

\begin{figure}
		\centering
		\includegraphics[width=\linewidth]{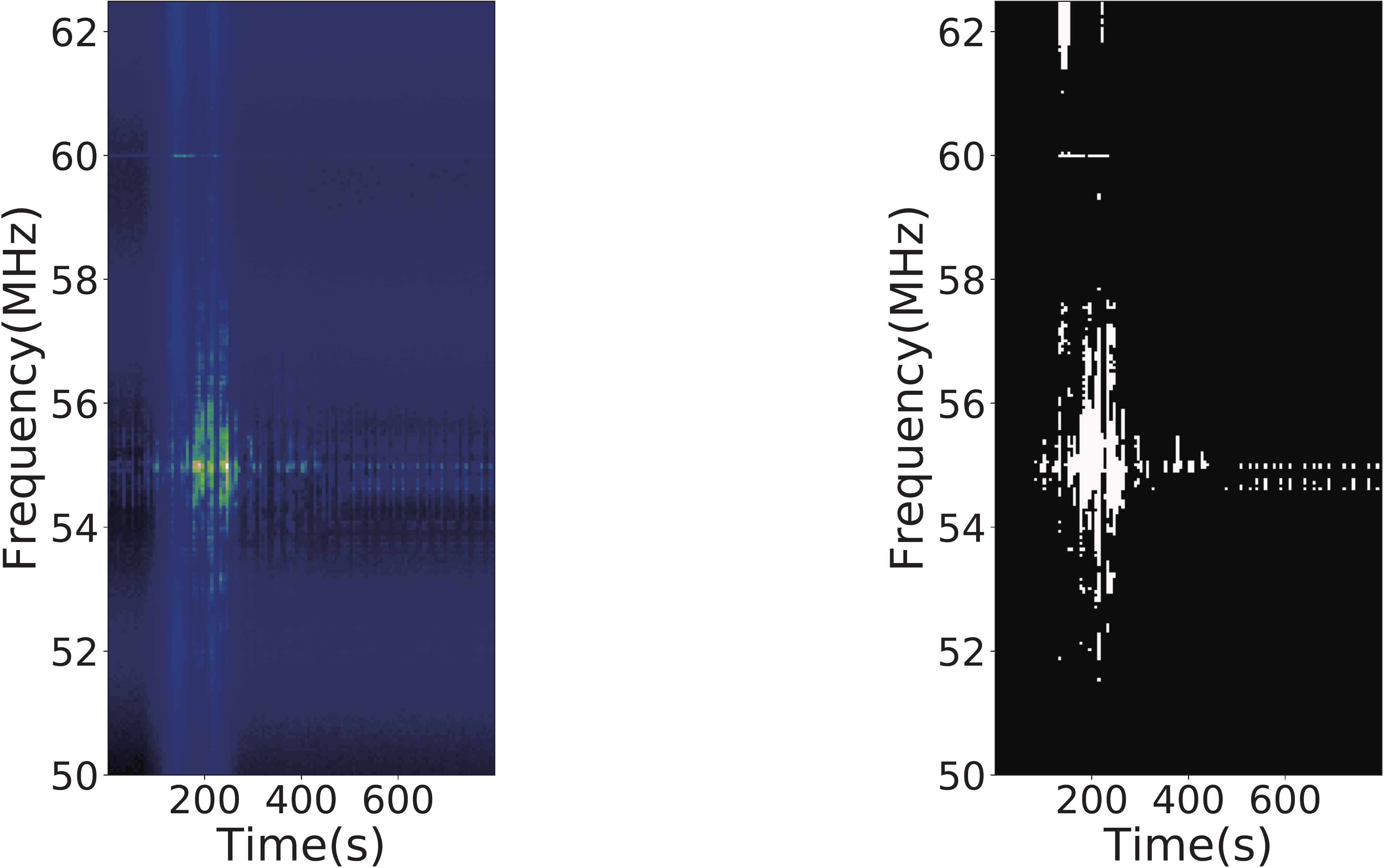}
		\caption{Detection result with data from Bleien Observatory. Most of the RFI has been identified, while there are falses positives and missing marks on the edge of the detected areas.}
		\label{Fig_4_bleien_data_rfi-net}
	\end{figure}

	The suitability of RFI-Net for practical applications has been evaluated with experiments on real data from the FAST telescope, as well as open access data from the Bleien Observatory. Fig.~\ref{Fig_4_real_data_rfi-net} shows the results obtained with the FAST dataset. Most of the RFI has been correctly identified with fine detection granularities, although very few instances have been missed. Results like this can be used as RFI masks without further processing. In addition, in Fig.~\ref{Fig_4_bleien_data_rfi-net}, RFI-Net also shows competitive performance. All evident RFI has been successfully flagged, although some insignificant ones with flux levels similar to background have missed detection.
	
	Since compared with observed data, our simulated data are simpler and have been more accurately marked, the accuracy of detection obtained with simulated data is higher than with real data. Thus, labelling observations more precisely would lead to further improvements in our results.
	
	\subsection{Accuracy with a smaller training dataset}
	\label{subsec_5.4_Accuracy on less training data}
	
\begin{figure}
		\centering
		\includegraphics[width=\linewidth]{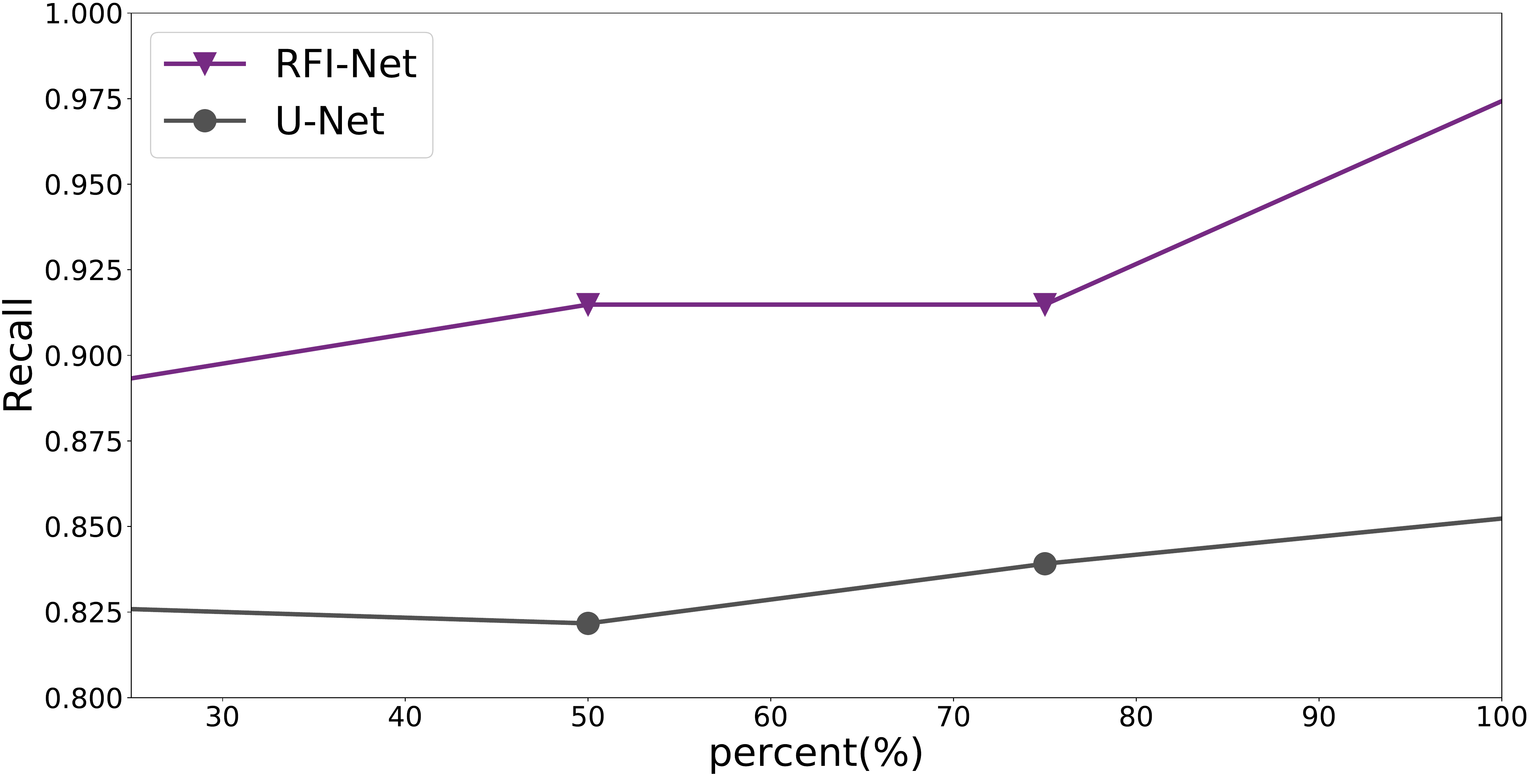}
		\caption{Recall of two models trained with different portions of the complete training dataset. It can be seen that with different sized dataset, the scores of RFI-Net are always higher than those of the U-Net, yielding more stable results.}
		\label{Fig_46}
	\end{figure}

\begin{figure}
		\centering
		\includegraphics[width=\linewidth]{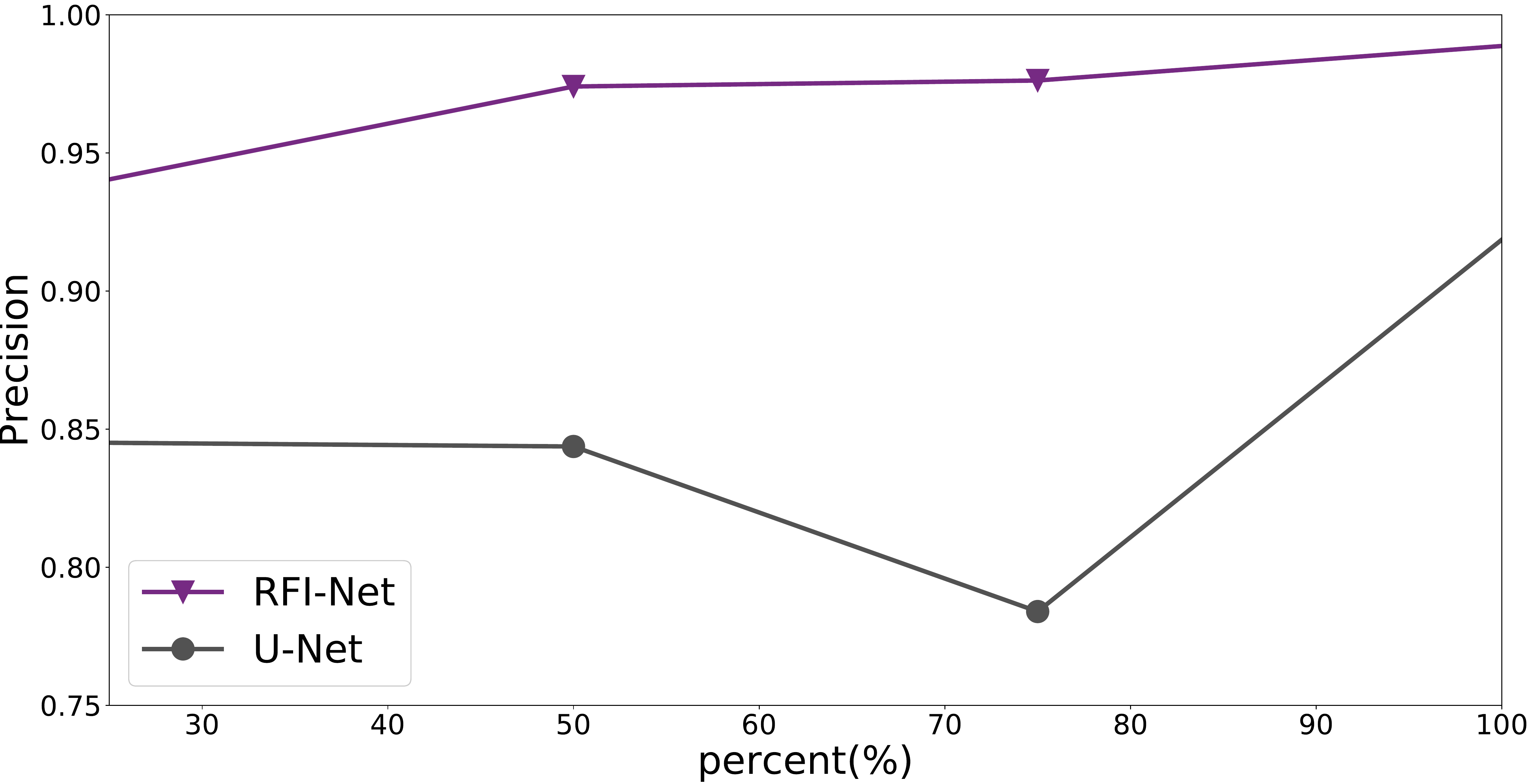}
		\caption{Precision scores of the two models with different portions of the complete training dataset. The performance of U-Net declines as the amount of training data reduced from 100\% to 75\%, while the RFI-Net model remains more stable.}
		\label{Fig_47}
	\end{figure}

\begin{figure}
		\centering
		\includegraphics[width=\linewidth]{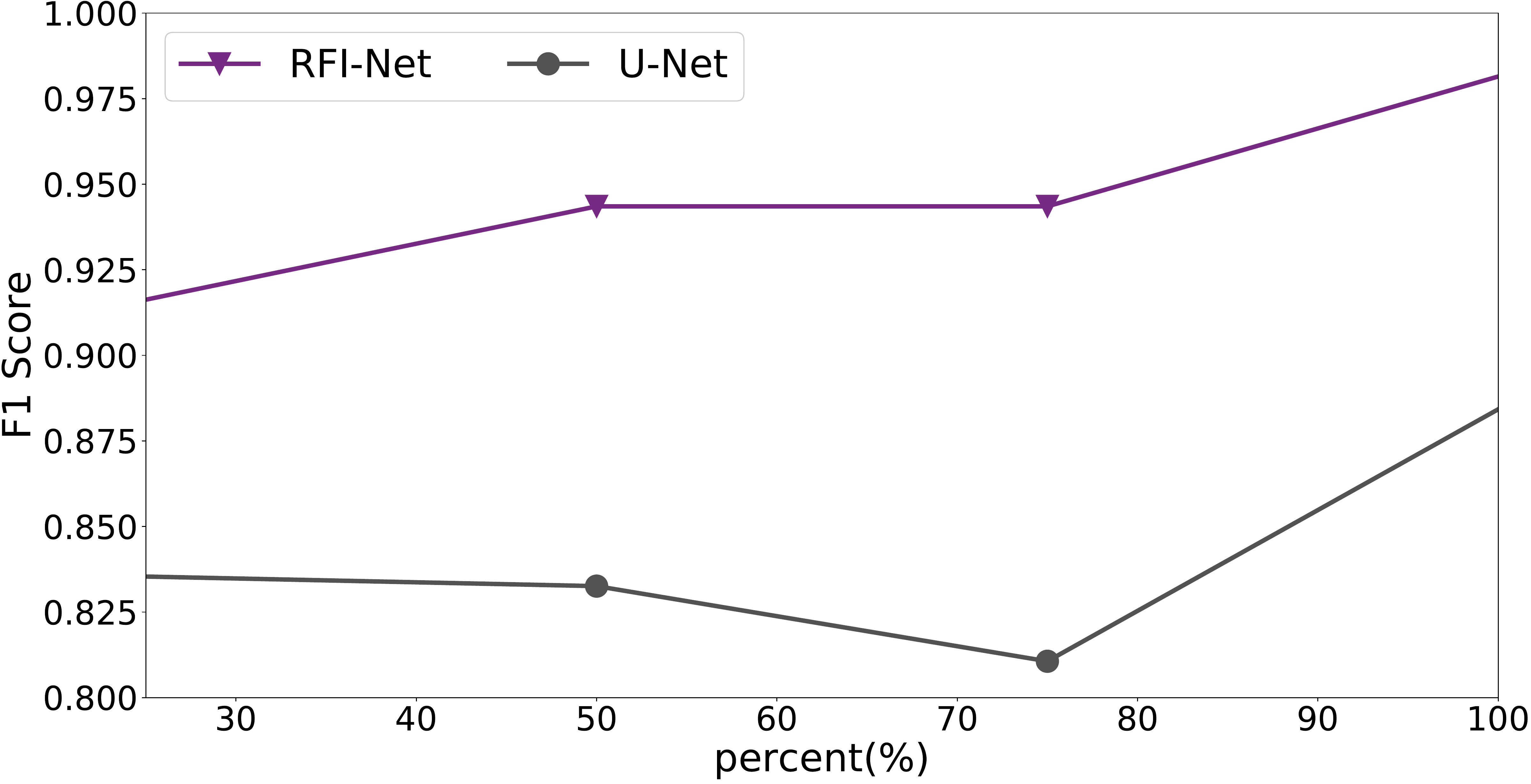}
		\caption{The overall performance of the two models, trained on different ratios of the whole training data, is judged by the F1 score. The result is similar to the precision. U-Net shows larger fluctuations, while RFI-Net is more stable, yielding higher scores.}
		\label{Fig_48}
	\end{figure}

	For smaller training dataset, Figures~\ref{Fig_46}, \ref{Fig_47}, and ~\ref{Fig_48} compare the results of our model with the original U-Net using all three performance metrics. These results confirm the outstanding performance of RFI-Net. It can be seen from Figure~\ref{Fig_46} that despite the volume of training data, the recalls of RFI-Net are higher than those of U-Net. When using all of the training data, the recall of RFI-Net is $97.5\%$, compared to $85\%$ of U-Net. While for $25\%$ of the dataset, although the difference between two models is smaller, the recall of RFI-Net ($89\%$) is still higher than that of the U-Net ($82\%$). It indicates that RFI-Net can make more comprehensive detections, with very few instances of RFI remaining unidentified. From the perspective of data volumes, a smaller amount of training data does affect the performance of the model in terms of completeness. The recall of RFI-Net has shown a significant decrease with $75\% - 100\%$ training dataset, in contrast to a more gradual decline of the original U-Net. However, with less than $75\%$ of original data, the amount of data has a lower impact on the recall of our model.
	
	Fig.~\ref{Fig_47} shows the precision of the detections with different training data volumes. It can be seen that the amount of training data had a more significant impact on the original U-Net than RFI-Net this time. The accuracy of U-Net experienced a decline when the data volume was reduced to 75\%, but then it restored from $77\%$ to $\sim 85\%$ as the data volume continued to be reduced. The performance of RFI-Net, on the other hand, is much more stable even without enough data for adequate training. The precision of our model shows a decrease with less than 50\% training data, which can lead to less accurate RFI detections.
	
	The F1 score, which is the reciprocal average (harmonic mean) of precision and recall, can be used to demonstrate the overall performance of each model, as shown in Fig.~\ref{Fig_48}. Still, the volume of training data can influence the performances of both models. However, RFI-Net delivered a much more stable curve, with a gradual decrease in the F1 score. In addition, RFI-Net maintains a higher score regardless of data volume. In contrast, U-Net performed less optimally, with a more consistent decline with the size of dataset.
	
	\subsection{Impacts of overfitting}
	\label{subsec_5.5_Impact on overfitting}
\begin{figure}
		\centering
		\includegraphics[width=\linewidth]{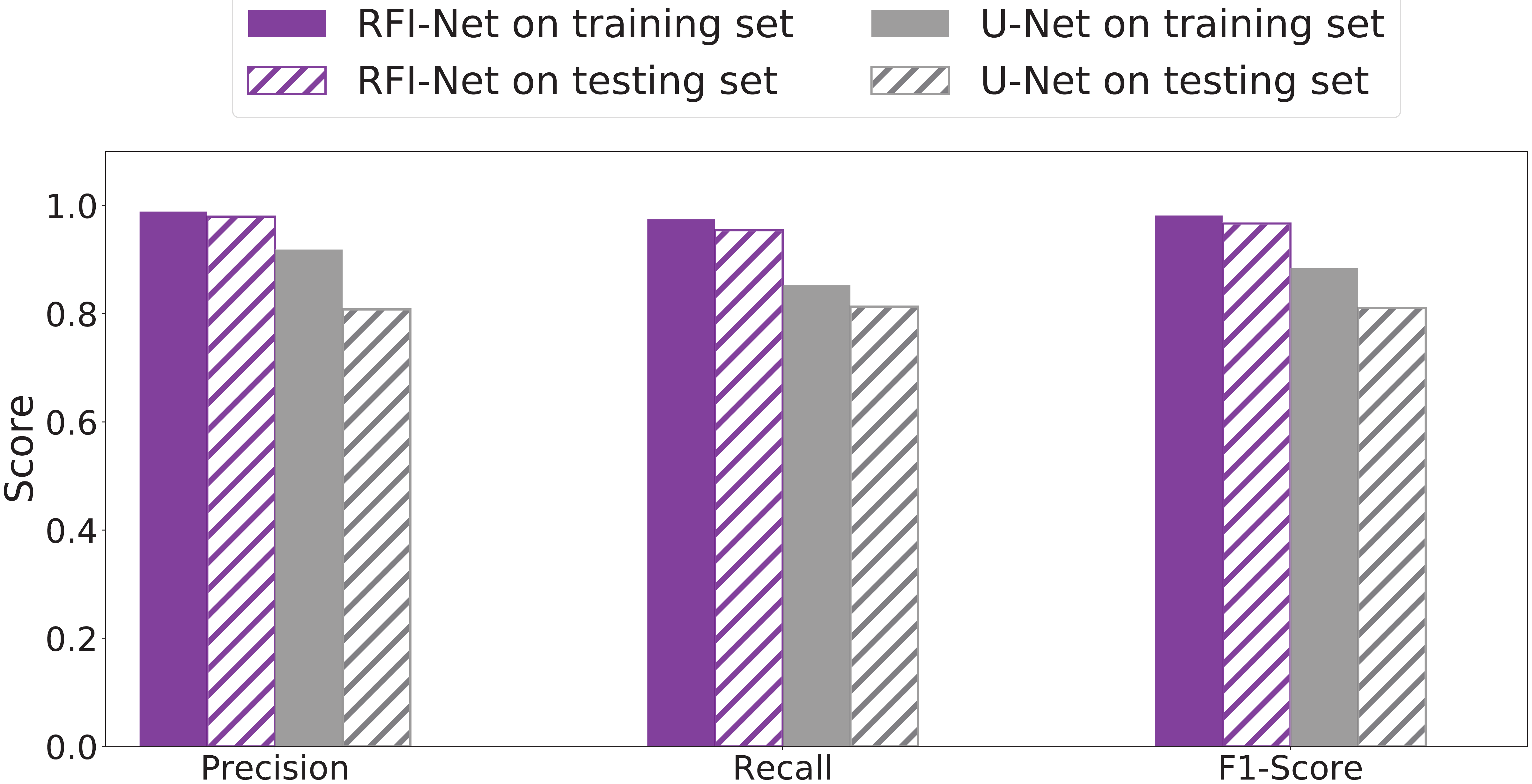}
		\caption{Results of RFI-Net on the overfitting problem. The difference between training and testing dataset with RFI-Net is much smaller than that of U-Net, which means overfitting can be further reduced by RFI-Net.}
		\label{Fig_4_result_on_overfitting_problem}
	\end{figure}

	Relatively lower accuracy acquired during practical applications compared with model training process is called the overfitting problem. Fig.~\ref{Fig_4_result_on_overfitting_problem} compares the performance of RFI-Net and U-Net with our training and testing datasets, respectively. Here the testing data comes from the same source as our training data, with a volume $\sim 10$\% of the latter. It can be seen that RFI-Net exhibits less degradation in the gradient when transferred from training set to testing set, whereas U-Net has degraded considerably.
	
	The detailed test results are listed in Table~\ref{Tab_4_result_on_overfitting_problem}. It is clearly shown that RFI-Net scored similarly on all indicators with both training and testing dataset. By comparison, U-Net only obtained much lower scores with our test data. However, it should be noted that such results are based upon simulated data. Although RFI-Net exhibits a better consistency with different datasets, it is possible that such consistency may degrade when dealing with more complicated observed data, and further investigations on this issue have been planned.

	\begin{table}
		\centering
		\caption{Detailed scores on overfitting problem with three performance metrics. It can be seen that our RFI-Net has achieved a more consistent performance.}
		\label{Tab_4_result_on_overfitting_problem}
		\begin{tabular*}{\linewidth}{@{\extracolsep{\fill}}ccccccc}
			\hline  
			&\multicolumn{2}{c}{Precision} &\multicolumn{2}{c}{Recall} &\multicolumn{2}{c}{F1 Score}\\
			& RFI-Net & U-Net & RFI-Net & U-Net & RFI-Net & U-Net\\
			\hline  
			training set & 98.87 & 91.86 & 97.42 & 85.23 & 98.14 & 88.43\\
			testing set & 97.93 & 80.78 & 95.43 & 81.30 & 96.67 & 81.04\\
			\hline
		\end{tabular*}
	\end{table}
	
	\subsection{Training speed of RFI-Net}
	\label{subsec_5.5_RFI-Net on training speed}
	
\begin{figure}
		\centering
		\includegraphics[width=\linewidth]{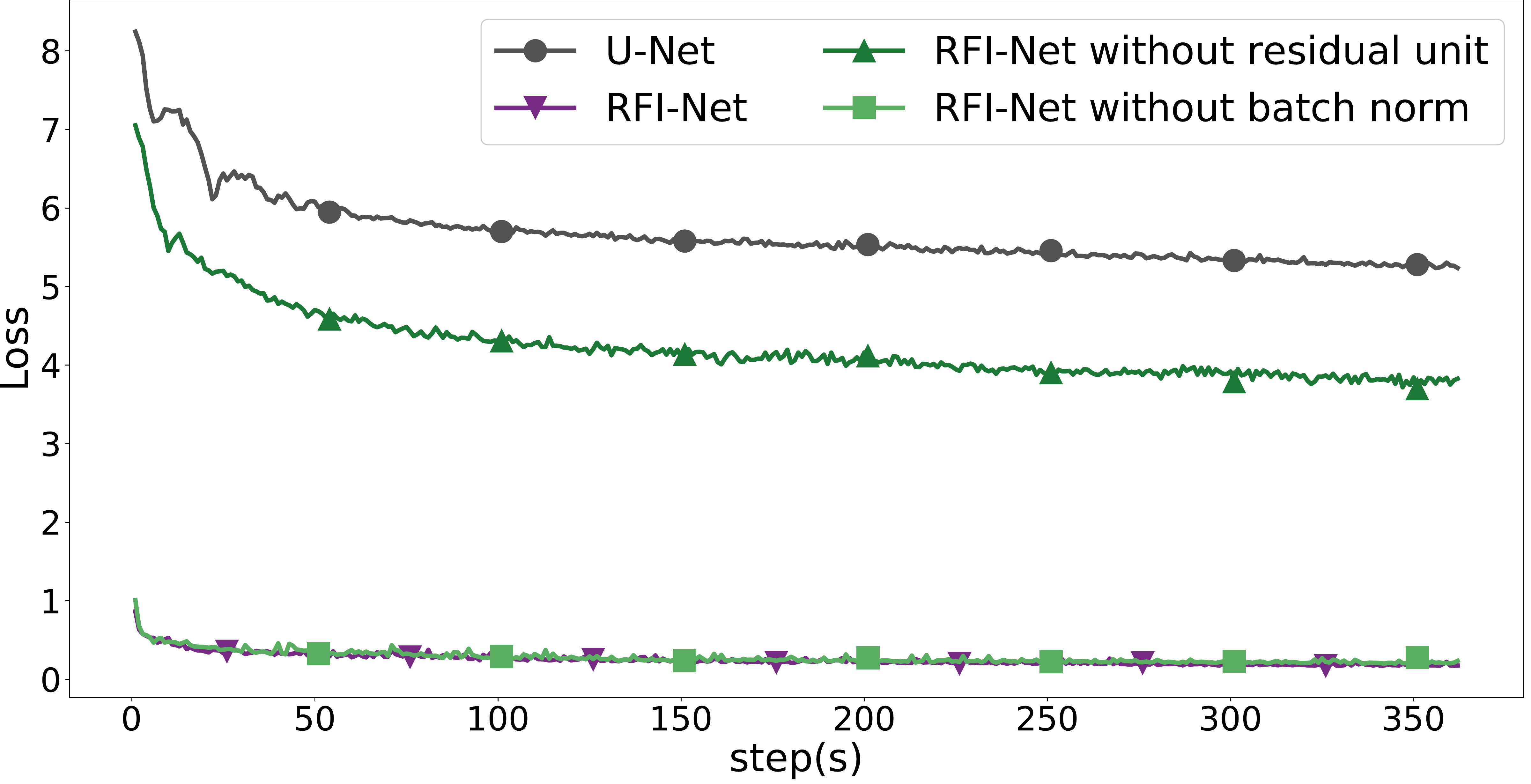}
		\caption{Loss curves produced by four configurations. The loss decreases as the number of steps increases. RFI-Net with and without batch normalisation (blue and brown, respectively) are both located at the lower part of the figure, since they converge early into straight lines. The losses of U-Net and RFI-Net without residual units (green and magenta) are so large, that log scale has been applied, in order to show the corresponding curves with residual units in the same graph.}
		\label{Fig_4_four_loss_in_trainning}
	\end{figure}

	Fig.~\ref{Fig_4_four_loss_in_trainning} shows the gradient descents of four configurations during training. Since the RFI-Net with residual units converges relatively fast, such models both with or without batch normalisation appear at the bottom of the graph. Compared with the original U-Net, as well as RFI-Net without residual units, it can be seen that the residual units can reduce the loss significantly. In addition, such effects take place earlier than that of the original U-Net, because of earlier stabilisation. Thus, with their presence, the residual units can accelerate the gradient descent effectively.
	
\begin{figure}
		\centering
		\includegraphics[width=\linewidth]{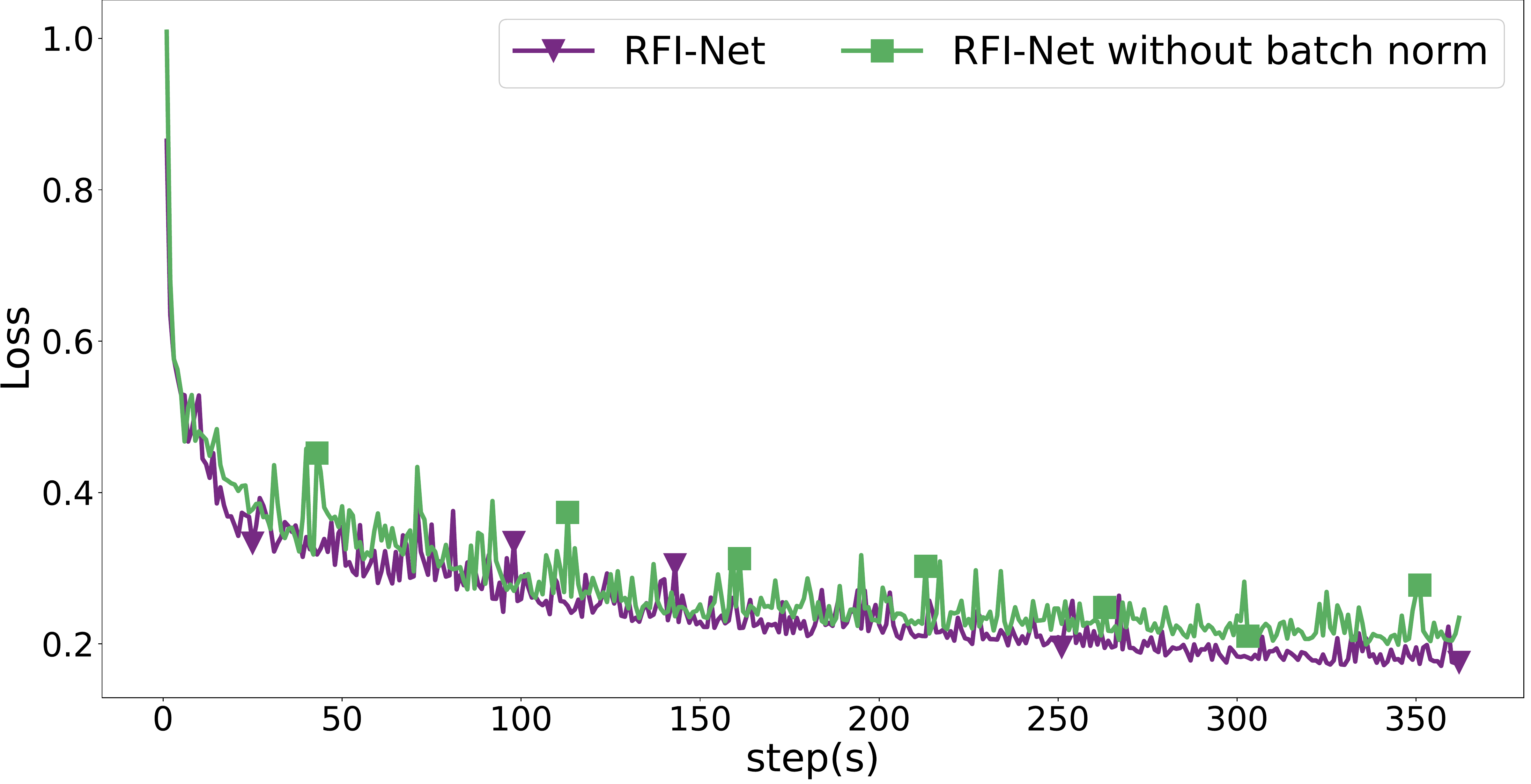}
		\caption{Zoom in view of the lower part of Fig.~\ref{Fig_4_four_loss_in_trainning}, showing loss curves of RFI-Net with and without batch normalisation. It can be seen that the introduction of BN can help to smooth and accelerate the process on a lesser (and insignificant) scale.}
		\label{Fig_4_two_loss_in_trainning}
	\end{figure}

	Fig.~\ref{Fig_4_two_loss_in_trainning} provides a zoomed-in version of the lower section of Fig.~\ref{Fig_4_four_loss_in_trainning}, showing the losses of RFI-Net with and without batch normalisation. The application of batch normalisation further improves the acceleration by three orders of magnitude. Thus, although the application of batch normalisation does not introduce much differences, even with the help of residual units, it still can facilitate the training process with faster acceleration.
	
\begin{figure}
		\centering
		\includegraphics[width=\linewidth]{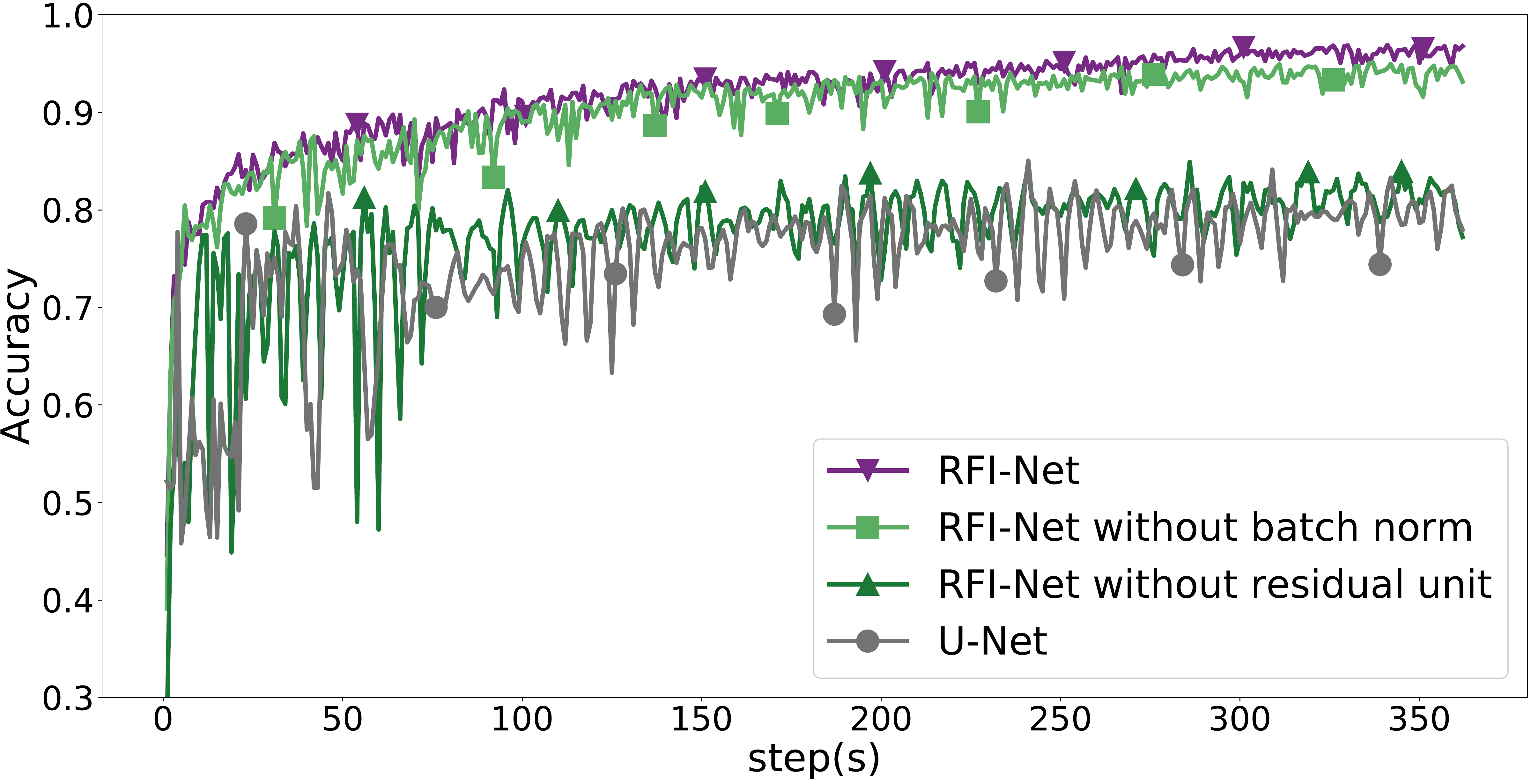}
		\caption{Accuracy of the training process. With the help of residual units, the RFI-Net can reach a stable state faster with less disturbances. The BN contributes a smaller portion to model stability and accuracy.}
		\label{Fig_4_four_accuracy_in_trainning}
	\end{figure}

	Fig.~\ref{Fig_4_four_accuracy_in_trainning} presents the training rate in terms of accuracy, and shows that RFI-Net can stabilise itself faster and more smoothly. RFI-Net with or without batch normalisation shows similar performance, although the latter model exhibits minor fluctuations and a slight decline in accuracy. By contrast, the case of RFI-Net without residual units shows off more variabilities, as well as much lower accuracy. Also, the performance of U-Net is similarly poor, with continued oscillations even when the accuracy has been stabilised.
	
	\section{Conclusion}
	\label{sec_6_Conclusion}
	Flagging radio frequency interference robustly is a challenging task. The high sensitivity and large data rate of the FAST telescope make RFI flagging and avoiding false positive a challenge. Although several studies have attempted to solve this problem, they either exhibit low-precision detections, or require a large amount of manual interventions, leading to a less-effective data reduction, and thus are unsuitable for FAST.
	
	We proposed the RFI-Net model in this paper, and have demonstrated that, compared with U-Net, KNN or SumThreshold methods, the performance of RFI-Net is superior in terms of RFI identification, making more comprehensive detections with higher accuracies, less errors, finer edges, as well as lower false positives possible. Our test results show no need for further manual inspections, thereby improving the efficiency of the data reduction pipeline. Our evaluations with a smaller training set have demonstrated the ability of RFI-Net to maintain its performance, with time and effort required for data preparation greatly reduced, overfitting minimised, and the training process accelerated.

	\section*{Acknowledgments}
	This work is supported by the Joint Research Fund in Astronomy (U1731125, U1731243) under a cooperative agreement between the National Natural Science Foundation of China (NSFC) and Chinese Academy of Sciences. BZ is also supported by NSFC grant 11903056, as well as the Open Project Program of the Key Laboratory of FAST, NAOC, Chinese Academy of Sciences.
	
	\section*{Conflicts of interest}
	The authors declare no conflict of interest.

	
	\bibliographystyle{mnras}
	\bibliography{Deep_residual_detection_of_Radio_Frequency_Interference_for_FAST}

\begin{thebibliography}{}
\makeatletter
\relax
\def\mn@urlcharsother{\let\do\@makeother \do\$\do\&\do\#\do\^\do\_\do\%\do\~}
\def\mn@doi{\begingroup\mn@urlcharsother \@ifnextchar [ {\mn@doi@}
  {\mn@doi@[]}}
\def\mn@doi@[#1]#2{\def\@tempa{#1}\ifx\@tempa\@empty \href
  {http://dx.doi.org/#2} {doi:#2}\else \href {http://dx.doi.org/#2} {#1}\fi
  \endgroup}
\def\mn@eprint#1#2{\mn@eprint@#1:#2::\@nil}
\def\mn@eprint@arXiv#1{\href {http://arxiv.org/abs/#1} {{\tt arXiv:#1}}}
\def\mn@eprint@dblp#1{\href {http://dblp.uni-trier.de/rec/bibtex/#1.xml}
  {dblp:#1}}
\def\mn@eprint@#1:#2:#3:#4\@nil{\def\@tempa {#1}\def\@tempb {#2}\def\@tempc
  {#3}\ifx \@tempc \@empty \let \@tempc \@tempb \let \@tempb \@tempa \fi \ifx
  \@tempb \@empty \def\@tempb {arXiv}\fi \@ifundefined
  {mn@eprint@\@tempb}{\@tempb:\@tempc}{\expandafter \expandafter \csname
  mn@eprint@\@tempb\endcsname \expandafter{\@tempc}}}

\bibitem[\protect\citeauthoryear{{Aguirre}, {Pichara}  \& {Becker}}{{Aguirre}
  et~al.}{2018}]{2019MNRAS.482.5078A}
{Aguirre} C.,  {Pichara} K.,   {Becker} I.,  2018, \mn@doi [Monthly Notices of
  the Royal Astronomical Society] {10.1093/mnras/sty2836}, 482, 5078

\bibitem[\protect\citeauthoryear{{Akeret}, {Seehars}, {Chang}, {Monstein},
  {Amara}  \& {Refregier}}{{Akeret} et~al.}{2017a}]{2017A&C....18....8A}
{Akeret} J.,  {Seehars} S.,  {Chang} C.,  {Monstein} C.,  {Amara} A.,
  {Refregier} A.,  2017a, \mn@doi [Astronomy and Computing]
  {10.1016/j.ascom.2016.11.001}, \href
  {https://ui.adsabs.harvard.edu/#abs/2017A&C....18....8A} {18, 8}

\bibitem[\protect\citeauthoryear{{Akeret}, {Chang}, {Lucchi}  \&
  {Refregier}}{{Akeret} et~al.}{2017b}]{2017A&C....18...35A}
{Akeret} J.,  {Chang} C.,  {Lucchi} A.,   {Refregier} A.,  2017b, \mn@doi
  [Astronomy and Computing] {10.1016/j.ascom.2017.01.002}, \href
  {https://ui.adsabs.harvard.edu/\#abs/2017A&C....18...35A} {18, 35}

\bibitem[\protect\citeauthoryear{An, Chen, MOHAN  \& Lao}{An
  et~al.}{2017}]{AnTao2017}
An T.,  Chen X.,  MOHAN P.,   Lao B.,  2017, \mn@doi [Acta Astronomica Sinica
  (in Chinese)] {10.15940/j.cnki.0001-5245.2017.05.002}, 58, 18

\bibitem[\protect\citeauthoryear{Arsalan, Kim, Lee, Owais  \& Park}{Arsalan
  et~al.}{2019}]{ARSALAN2019217}
Arsalan M.,  Kim D.~S.,  Lee M.~B.,  Owais M.,   Park K.~R.,  2019, \mn@doi
  [Expert Systems with Applications]
  {https://doi.org/10.1016/j.eswa.2019.01.010}, 122, 217

\bibitem[\protect\citeauthoryear{{Baan}, {Fridman}  \& {Millenaar}}{{Baan}
  et~al.}{2004}]{2004AJ....128..933B}
{Baan} W.~A.,  {Fridman} P.~A.,   {Millenaar} R.~P.,  2004, \mn@doi [The
  Astronomical Journal] {10.1086/422350}, \href
  {https://ui.adsabs.harvard.edu/\#abs/2004AJ....128..933B} {128, 933}

\bibitem[\protect\citeauthoryear{Boureau, Ponce  \& LeCun}{Boureau
  et~al.}{2010}]{Boureau:2010:TAF:3104322.3104338}
Boureau Y.-L.,  Ponce J.,   LeCun Y.,  2010, in Proceedings of the 27th
  International Conference on International Conference on Machine Learning.
  ICML'10.
Omnipress, USA, pp 111--118, \url
  {http://dl.acm.org/citation.cfm?id=3104322.3104338}

\bibitem[\protect\citeauthoryear{Briggs, Sorar, Kraan-Korteweg  \& van
  Driel}{Briggs et~al.}{1997}]{1997PASA...14...37B}
Briggs F.,  Sorar E.,  Kraan-Korteweg R.,   van Driel W.,  1997, \mn@doi
  [Publications of the Astronomical Society of Australia] {10.1071/AS97037},
  14, 37

\bibitem[\protect\citeauthoryear{Burd, Mannheim, M{\"{a}}rz, Ringholz, Kappes
  \& Kadler}{Burd et~al.}{2018}]{2018AN....339..358B}
Burd P.,  Mannheim K.,  M{\"{a}}rz T.,  Ringholz J.,  Kappes A.,   Kadler M.,
  2018, \mn@doi [Astronomische Nachrichten] {10.1002/asna.201813505}, 339, 358

\bibitem[\protect\citeauthoryear{Czech, Mishra  \& Inggs}{Czech
  et~al.}{2018}]{Czech2018}
Czech D.,  Mishra A.,   Inggs M.,  2018, \mn@doi [Astronomy and Computing]
  {10.1016/J.ASCOM.2018.07.002}, 25, 52

\bibitem[\protect\citeauthoryear{Davis \& Goadrich}{Davis \&
  Goadrich}{2006}]{davis2006relationship}
Davis J.,  Goadrich M.,  2006, in Proceedings of the 23rd international
  conference on Machine learning. pp 233--240

\bibitem[\protect\citeauthoryear{Drozdzal, Vorontsov, Chartrand, Kadoury  \&
  Pal}{Drozdzal et~al.}{2016}]{Drozdzal2016TheIO}
Drozdzal M.,  Vorontsov E.,  Chartrand G.,  Kadoury S.,   Pal C.~J.,  2016, in
  LABELS/DLMIA@MICCAI.

\bibitem[\protect\citeauthoryear{ETHZurich}{ETHZurich}{2016}]{BleienObservatoryData}
ETHZurich 2016, Raw data from the Bleien Observatory, \url
  {https://people.phys.ethz.ch/~ipa/cosmo/bgs_example_data/}

\bibitem[\protect\citeauthoryear{Evans, Owda, Crockett  \& Vilas}{Evans
  et~al.}{2019}]{EVANS2019353}
Evans L.,  Owda M.,  Crockett K.,   Vilas A.,  2019, \mn@doi [Expert Systems
  with Applications] {https://doi.org/10.1016/j.eswa.2019.03.019}, 127, 353

\bibitem[\protect\citeauthoryear{G{\'{o}}mez-R{\'{i}}os, Tabik, Luengo,
  Shihavuddin, Krawczyk  \& Herrera}{G{\'{o}}mez-R{\'{i}}os
  et~al.}{2019}]{GOMEZRIOS2019315}
G{\'{o}}mez-R{\'{i}}os A.,  Tabik S.,  Luengo J.,  Shihavuddin A. S.~M.,
  Krawczyk B.,   Herrera F.,  2019, \mn@doi [Expert Systems with Applications]
  {https://doi.org/10.1016/j.eswa.2018.10.010}, 118, 315

\bibitem[\protect\citeauthoryear{{Guizhou Provincial People's
  Government}}{{Guizhou Provincial People's Government}}{2019}]{Government2019}
{Guizhou Provincial People's Government} 2019, Regulations for protection of
  electromagnetic wave quiet zone of 500 meter aperture spherical radio
  telescope in guizhou province, \url
  {http://www.guizhou.gov.cn/zwgk/zcfg/szfwj_8191/szfl_8192/201901/t20190121_2222872.html}

\bibitem[\protect\citeauthoryear{Guo, Wang, Bell, Bi  \& Greer}{Guo
  et~al.}{2003}]{10.1007/978-3-540-39964-3_62}
Guo G.,  Wang H.,  Bell D.,  Bi Y.,   Greer K.,  2003, in Meersman R.,  Tari
  Z.,   Schmidt D.~C.,  eds, On The Move to Meaningful Internet Systems 2003:
  CoopIS, DOA, and ODBASE. Springer Berlin Heidelberg, Berlin, Heidelberg, pp
  986--996

\bibitem[\protect\citeauthoryear{{Haiyan Zhang}, Nan, {Bo Peng}, {Yuebing Xia},
  Jin, Li, {Xiaonian Zheng}  \& {Long Gao}}{{Haiyan Zhang}
  et~al.}{2013}]{2013APS........14}
{Haiyan Zhang} Nan R.,  {Bo Peng} {Yuebing Xia} Jin C.,  Li J.,  {Xiaonian
  Zheng}  {Long Gao} 2013, in 2013 Asia-Pacific Symposium on Electromagnetic
  Compatibility (APEMC). pp~1--3, \mn@doi{10.1109/APEMC.2013.7360597}

\bibitem[\protect\citeauthoryear{He \& Sun}{He \&
  Sun}{2015}]{2014arXiv1412.1710H}
He K.,  Sun J.,  2015, in Proceedings of the IEEE conference on computer vision
  and pattern recognition. pp 5353--5360

\bibitem[\protect\citeauthoryear{He, Zhang, Ren  \& Sun}{He
  et~al.}{2016a}]{he2016identity}
He K.,  Zhang X.,  Ren S.,   Sun J.,  2016a, in European conference on computer
  vision. pp 630--645

\bibitem[\protect\citeauthoryear{{He}, {Zhang}, {Ren}  \& {Sun}}{{He}
  et~al.}{2016b}]{2015arXiv151203385H}
{He} K.,  {Zhang} X.,  {Ren} S.,   {Sun} J.,  2016b, in Proceedings of the IEEE
  conference on computer vision and pattern recognition. pp 770--778
  (\mn@eprint {} {1512.03385}), \mn@doi{10.1007/s11042-017-4440-4}

\bibitem[\protect\citeauthoryear{Indermuehle, Harvey-Smith, Wilson  \&
  Chow}{Indermuehle et~al.}{2016}]{7833529}
Indermuehle B.~T.,  Harvey-Smith L.,  Wilson C.,   Chow K.,  2016, in 2016
  Radio Frequency Interference (RFI). pp 43--48,
  \mn@doi{10.1109/RFINT.2016.7833529}

\bibitem[\protect\citeauthoryear{{Ioffe} \& {Szegedy}}{{Ioffe} \&
  {Szegedy}}{2015}]{2015arXiv150203167I}
{Ioffe} S.,  {Szegedy} C.,  2015, arXiv e-prints, \href
  {https://ui.adsabs.harvard.edu/\#abs/2015arXiv150203167I} {p.
  arXiv:1502.03167}

\bibitem[\protect\citeauthoryear{Kingma \& Ba}{Kingma \& Ba}{2015}]{Kingma2014}
Kingma D.~P.,  Ba J.,  2015, \mn@doi [International Conference on Learning
  Representations]
  {http://doi.acm.org.ezproxy.lib.ucf.edu/10.1145/1830483.1830503}, pp 1--15

\bibitem[\protect\citeauthoryear{Krizhevsky, Sutskever  \& Hinton}{Krizhevsky
  et~al.}{2012}]{2012:ICD:2999134.2999257}
Krizhevsky A.,  Sutskever I.,   Hinton G.~E.,  2012, in Proceedings of the 25th
  International Conference on Neural Information Processing Systems - Volume 1.
  NIPS'12.
Curran Associates Inc., USA, pp 1097--1105, \url
  {http://dl.acm.org/citation.cfm?id=2999134.2999257}

\bibitem[\protect\citeauthoryear{{Lv}, {Jiang}  \& {Li}}{{Lv}
  et~al.}{2017}]{2017arXiv170303633L}
{Lv} K.,  {Jiang} S.,   {Li} J.,  2017, arXiv e-prints, \href
  {https://ui.adsabs.harvard.edu/abs/2017arXiv170303633L} {p. arXiv:1703.03633}

\bibitem[\protect\citeauthoryear{{Mosiane}, {Oozeer}, {Aniyan}  \&
  {Bassett}}{{Mosiane} et~al.}{2017}]{2017MS&E..198a2012M}
{Mosiane} O.,  {Oozeer} N.,  {Aniyan} A.,   {Bassett} B.~A.,  2017, in
  Materials Science and Engineering Conference Series. p. 012012,
  \mn@doi{10.1088/1757-899X/198/1/012012}

\bibitem[\protect\citeauthoryear{Nan et~al.,}{Nan et~al.}{2011}]{2011Nan}
Nan R.,  et~al., 2011, International Journal of Modern Physics D, 20, 989

\bibitem[\protect\citeauthoryear{Offringa, de Bruyn, Zaroubi  \&
  Biehl}{Offringa et~al.}{2010a}]{Offringa2010a}
Offringa A.~R.,  de Bruyn A.~G.,  Zaroubi S.,   Biehl M.,  2010a,
  Instrumentation and Methods for Astrophysics

\bibitem[\protect\citeauthoryear{{Offringa}, {de Bruyn}, {Biehl}, {Zaroubi},
  {Bernardi}  \& {Pandey}}{{Offringa} et~al.}{2010b}]{2010MNRAS.405..155O}
{Offringa} A.~R.,  {de Bruyn} A.~G.,  {Biehl} M.,  {Zaroubi} S.,  {Bernardi}
  G.,   {Pandey} V.~N.,  2010b, \mn@doi [Monthly Notices of the Royal
  Astronomical Society] {10.1111/j.1365-2966.2010.16471.x}, \href
  {https://ui.adsabs.harvard.edu/#abs/2010MNRAS.405..155O} {405, 155}

\bibitem[\protect\citeauthoryear{Pedregosa et~al.,}{Pedregosa
  et~al.}{2011}]{2011scikitlearn11}
Pedregosa F.,  et~al., 2011, Journal of Machine Learning Research, 12, 2825

\bibitem[\protect\citeauthoryear{Popping \& Braun}{Popping \&
  Braun}{2008}]{2008A&A...479..903P}
Popping A.,  Braun R.,  2008, \mn@doi [A{\&}A] {10.1051/0004-6361:20079122},
  479, 903

\bibitem[\protect\citeauthoryear{Ronneberger, Fischer  \& Brox}{Ronneberger
  et~al.}{2015}]{2015arXiv150504597R}
Ronneberger O.,  Fischer P.,   Brox T.,  2015, in Navab N.,  Hornegger J.,
  Wells W.~M.,   Frangi A.~F.,  eds, Medical Image Computing and
  Computer-Assisted Intervention -- MICCAI 2015. Springer International
  Publishing, Cham, pp 234--241

\bibitem[\protect\citeauthoryear{{Ruder}}{{Ruder}}{2016}]{2016arXiv160904747R}
{Ruder} S.,  2016, arXiv e-prints, \href
  {https://ui.adsabs.harvard.edu/abs/2016arXiv160904747R/abstract/} {p.
  arXiv:1609.04747}

\bibitem[\protect\citeauthoryear{Srivastava, Hinton, Krizhevsky, Sutskever  \&
  Salakhutdinov}{Srivastava et~al.}{2014}]{JMLR:v15:srivastava14a}
Srivastava N.,  Hinton G.,  Krizhevsky A.,  Sutskever I.,   Salakhutdinov R.,
  2014, Journal of Machine Learning Research, 15, 1929

\bibitem[\protect\citeauthoryear{Wilson, Chow, Harvey-Smith, Indermuehle,
  Sokolowski  \& Wayth}{Wilson et~al.}{2016}]{7731554}
Wilson C.,  Chow K.,  Harvey-Smith L.,  Indermuehle B.,  Sokolowski M.,   Wayth
  R.,  2016, in 2016 International Conference on Electromagnetics in Advanced
  Applications (ICEAA). pp 922--923, \mn@doi{10.1109/ICEAA.2016.7731554}

\bibitem[\protect\citeauthoryear{Wolfaardt}{Wolfaardt}{2016}]{Wolfaardt2016}
Wolfaardt C.~J.,  2016, PhD thesis, Stellenbosch University

\bibitem[\protect\citeauthoryear{{Zhu} et~al.,}{{Zhu} et~al.}{2014}]{2014Zhu}
{Zhu} W.~W.,  et~al., 2014, \mn@doi [\apj] {10.1088/0004-637X/781/2/117}, \href
  {https://ui.adsabs.harvard.edu/abs/2014ApJ...781..117Z} {781, 117}

\makeatother
\end{thebibliography}

	\bsp	
	\label{lastpage}
\end{document}